\documentclass[12pt]{article}

\ifx\pdfoutput\undefined
\usepackage[dvips,bookmarks]{hyperref}
\else
\usepackage{hyperref}
\fi
\hypersetup{colorlinks=false,bookmarksopen,bookmarksnumbered,citecolor=blue,
   pdfstartview=FitH}

\usepackage{latexsym}
\usepackage{amssymb,amsfonts,amsmath}
\usepackage{graphicx}
\usepackage{indentfirst}
\usepackage{bbm}
\usepackage{amssymb}
\usepackage{verbatim}
\usepackage{amsmath, amsthm,amssymb}
\usepackage{mathrsfs}
\usepackage{hyperref}
\usepackage{amsfonts}
\usepackage{dsfont}
\usepackage{slashed, tensor}
\usepackage{booktabs}
\usepackage{graphicx}
\usepackage{mathrsfs}

\parskip=\medskipamount

\newcommand {\cA}{{\cal A}}
\newcommand {\cB}{{\cal B}}
\newcommand {\cC}{{\cal C}}
\newcommand {\cD}{{\cal D}}

\newcommand {\cH}{{\cal H}}

\newcommand {\cL}{{\cal L}}

\newcommand {\cN}{{\cal N}}
\newcommand {\cO}{{\cal O}}

\def\a{\alpha}
\def\b{\beta}

\def\d{\delta}

\def\g{\gamma}
\def\G{\Gamma}

\def\k{\kappa}
\def\l{\lambda}

\def\q{\theta}
\def\r{\rho}

\def\t{\tau}

\def\x{\xi}

\def\S{\Sigma}

\def\tr{{\rm tr}}
\def\Tr{{\rm Tr}}

\def\ri{{\rm i}}
\def\re{{\rm e}}

\newcommand{\ad}{{\dot{\alpha}}}                           
\newcommand{\ve}{\varepsilon}                            

\newcommand{\be}{\begin{equation}}
\newcommand{\ee}{\end{equation}}
\newcommand{\bea}{\begin{eqnarray}}
\newcommand{\eea}{\end{eqnarray}}
\newcommand{\non}{\nonumber}
\newcommand{\ba}{\begin{array}}
\newcommand{\ea}{\end{array}}

\def\double #1{#1{\hbox{\kern-2pt $#1$}}}

\newcommand{\sSU}{\mathsf{SU}}

\newcommand{\sU}{\mathsf{U}}

\newcommand{\bsubeq}{\begin{subequations}}
\newcommand{\esubeq}{\end{subequations}}

\newcommand{\eps}{{\ve}}

\newcommand{\rd}{\mathrm d}
\numberwithin{equation}{section}

\begin{document}

\begin{titlepage}
\begin{flushright}
August, 2017\\
\end{flushright}
\vspace{5mm}

\begin{center}
{\Large \bf Chiral  anomalies in six dimensions
\\[2mm] from harmonic superspace
}
\\
\end{center}

\begin{center}

{\bf
Sergei M. Kuzenko$^a$,
Joseph Novak$^b$
and Igor B. Samsonov$^c$
} \\
\vspace{5mm}

\footnotesize{
{\it ${}^a$ School of Physics and Astrophysics M013, The University of Western Australia,\\
35 Stirling Highway, Crawley, W.A. 6009, Australia\\
{\tt sergei.kuzenko@uwa.edu.au}

\vspace{2mm}
${}^{b}$ Max-Planck-Institut f\"ur Gravitationsphysik, Albert-Einstein-Institut,\\
Am M\"uhlenberg 1, D-14476 Golm, Germany. \\
{\tt joseph.novak@aei.mpg.de}

\vspace{2mm}
$^c$ Bogoliubov Laboratory of Theoretical Physics, JINR,\\
141980 Dubna, Moscow region, Russia}
\\
{\tt samsonov@theor.jinr.ru}}

\vspace{2mm}
~\\

\end{center}

\begin{abstract}
\baselineskip=14pt
We develop a superfield approach to compute chiral anomalies in
general $\cN=(1,0)$ supersymmetric gauge theories in six dimensions.
Within the harmonic-superspace formulation for these gauge theories, the anomalous
contributions to the effective action only come from matter and ghost
hypermultiplets. By studying the short-distance behaviour of the propagator for the hypermultiplet coupled to a
background vector multiplet,
we compute the covariant and consistent chiral anomalies.
We also provide a superform formulation for
the non-abelian anomalous current multiplet
 in general $\cN=(1,0)$ supersymmetric gauge theories.
\end{abstract}

\vfill

\vfill
\end{titlepage}

\newpage
\renewcommand{\thefootnote}{\arabic{footnote}}
\setcounter{footnote}{0}


\tableofcontents
\vspace{1cm}
\bigskip\hrule



\section{Introduction}

The general structure of chiral and gravitational anomalies in
gauge theories in diverse dimensions
was fully understood as long ago as
the mid-1980s \cite{WZ-cons,Fujikawa,Stora,Zumino,A-GW,ZuminoZee,Bardeen,Leutwyler}
(more complete lists of references can be found, e.g.,  in \cite{FS,BvN}).
In supersymmetric gauge theories, both chiral and gravitational anomalies (if present)
should be embedded into supermultiplets. It is somewhat surprising that not much
explicit information is available
about the anomaly supermultiplets in six and higher spacetime dimensions,
in contrast to four dimensions (4D).\footnote{Chiral anomalies for
general supersymmetric gauge theories in dimensions $D= 6,8$ and 10 were computed in \cite{TS}, and can be read off from the results in
\cite{A-GW,ZuminoZee,FK1,FK2,FK3},
however no discussion of anomaly supermultiplets was given in these publications.}
As is well known, the latter case
is characterised  by the absence of gravitational anomalies,
while chiral gauge anomalies cannot occur in
extended supersymmetry  ($\cN>1$). We recall that the fermions in general 4D $\cN=2$ supersymmetric gauge theories transform in non-chiral representations
of the gauge group.

The chiral anomalies in 4D $\cN=1$ supersymmetric gauge theories
have  been thoroughly  studied in numerous works, see
\cite{PS84,CL84,Nielsen84,GGS,GKM85,KS85,BPT1,BPT2,OMc,J,Maja,GGKPS} and
references therein.
These studies include both the formal aspects of supersymmetric gauge anomalies
as well as the powerful superfield techniques to compute such anomalies.
In particular, it was realised that the (abelian)
chiral anomaly can be viewed as a consistent deformation of the $\cN=1$ linear
multiplet $L=\bar L$. In the anomaly-free case, $L$ obeys the constraint
$\bar D^2 L=0$, which implies that the component field $[D_\a, \bar D_\ad] L|_{\q=0}$
is a conserved current \cite{FWZ}.
If a chiral anomaly is present, the conservation
equation is deformed to take the form
\bea
\bar {D}_{\dot\alpha}\bar{D}^{\dot\alpha} L \propto W^\alpha W_\alpha~,
\label{4Danomaly}
\eea
where $W_\alpha$ is the chiral gauge-invariant field strength
of a vector multiplet.

In the case of 6D $\cN=(1,0)$ supersymmetry, a conserved current belongs to
the linear multiplet $L^{ij}$ which is a real
$\sSU(2)$ iso-triplet constrained by $D^{(k}_\alpha L^{ij)} =0$ \cite{BSohnius}.
In the presence of a chiral anomaly, this conservation equation
turns into a  deformed one. It was shown in
our recent paper \cite{KNS} (see
also \cite{HSezgin}) that
the abelian chiral anomaly amounts to
the following deformation
\be
A^{ijk}_\alpha:={D}^{(k}_\alpha L^{ij)} \propto
\ri\,\varepsilon_{\alpha\beta\gamma\delta}W^{i\beta}W^{j\gamma}W^{k\delta}~,
\label{1}
\ee
where $W^{i\alpha}$ is the gauge-invariant field strength of a
vector multiplet.\footnote{The modern superfield formulation for 6D $\cN=(1,0)$
supersymmetric Yang-Mills theory
was developed in \cite{HST,Koller}
as a reformulation of the earlier $\sSU(2) $ non-covariant approach
 \cite{Siegel,Nilsson}.}
 In \cite{KNS} we also provided a nonlocal effective action
 which generates the anomaly \eqref{1}.
So far, however, even the abelian chiral anomaly
(\ref1) has never been
computed directly in superspace in spite of the recent
progress in applying the background covariant  supergraphs
to compute low-energy effective actions for  6D $\cN=(1,0)$
supersymmetric gauge theories \cite{BP,BP2,BP3,BP4,BP5,BP6}.
The present paper aims to
fill this gap by providing an explicit supergraph derivation of the {\it non}-abelian chiral
anomaly using the  6D $\cN=(1,0) $ harmonic-superspace setting \cite{HSWest,Zupnik86,Zupnik:1987vm}.

In  6D $\cN=(1,0) $ harmonic superspace, the non-abelian vector multiplet
can be described either
in terms of the analytic prepotential $V^{++}$ \cite{HSWest,Zupnik86,Zupnik:1987vm}
(which is introduced in complete analogy with the 4D $\cN=2$ case \cite{GIKOS})
or by means of an unconstrained harmonic superfield
$M^{--}$ defined by $V^{++} = ({D}^+)^4 M^{--}$ \cite{GIOS}.
We will refer to $M^{--}$ as the ``generalised Mezincescu prepotential''
due to the fact that $M^{--}$ contains the 6D analogue \cite{HST,Koller} of
Mezincescu's prepotential $M^{ij}$ \cite{Mezincescu}.\footnote{The prepotential
 $M^{--}$ contains the
conventional Mezincescu prepotential $M^{ij}$
as a leading Fourier coefficient in its harmonic expansion,
 $M^{--}(z,u)=M^{ij}(z)u^-_i u^-_j+\ldots$ As explained in \cite{GIOS},
a gauge condition may be chosen in which $M^{--}(z,u)=M^{ij}(z)u^-_i u^-_j$.
 }
In order to compute the chiral anomalies using supergraphs, one may
analyse the effective action $\Gamma=\Gamma[M^{--}]$
in a general 6D $\cN=(1,0)$ supersymmetric gauge theory.
Such a theory describes the pure supersymmetric Yang-Mills (SYM) theory coupled to a hypermultiplet transforming in some representation of the gauge group.
In this paper we argue that the non-abelian extension of (\ref1)
naturally originates  as a covariant chiral anomaly.
We also generalise Leutwyler's ideas \cite{Leutwyler}
to the 6D $\cN=(1,0)$ harmonic-superspace setting in order to construct a
consistent chiral anomaly from the covariant one.

This paper is organised as follows.
We first present a superform formulation of
the non-abelian anomalous current multiplet in general 6D $\cN=(1,0)$ supersymmetric gauge theories in section  \ref{Sec4}.
This construction allows for two possible forms of the
non-abelian chiral anomaly, which
reduce to the same expression in the abelian case. Explicit
supergraph computation presented in subsequent sections shows that
only one of these structures is realised as a part of effective action in (1,0) gauge
theories.

To compute chiral anomalies in general 6D $\cN=(1,0)$ supersymmetric gauge theories,
it suffices  to consider the hypermultiplet
model in the presence of  a background vector multiplet. Indeed, in pure
SYM theory the anomalous contributions  can come only from ghost superfields which are described by hypermultiplets according to
\cite{BFM,CommentsBFM,KM0}. Therefore, following the discussion of general
aspects of covariant and consistent anomalies in superspace given in section \ref{SecChirAnom},
in section \ref{Sec3} we
examine the short-distance behaviour of the hypermultiplet propagator in
the presence of a background vector
multiplet and argue that the non-abelian extension of (\ref1)
naturally arises from those terms in the propagator which involve
harmonic singularities. In the 4D $\cN=2$ and 5D $\cN=1$  cases, similar
terms in the hypermultiplet propagator give no
contributions to the effective action \cite{KM,K-5D}, in agreement with the fact
that all 4D $\cN=2$ supersymmetric gauge theories
have no chiral anomaly (which is also obviously true for the  5D $\cN=1$ theories).
This uncovers the role of harmonic singularities in the
context of chiral anomalies.

In the main body of the paper, we mostly
study the form of the chiral anomaly in the
formulation of the (1,0) gauge theory based on the generalised
Mezincescu prepotential $M^{--}$. However, the concluding section \ref{Summary}
discusses the issues of describing
the chiral anomaly in the formulation of gauge theory
with the analytic prepotential $V^{++}$. In particular, we propose a
consistent expression for this anomaly in the abelian case.

In this paper, we follow the 6D superspace notation and conventions
given in \cite{LT-M} and employed in our recent paper \cite{KNS}.
In Appendix A we review the basic aspects of the 6D supersymmetric gauge theories
in $\cN=(1,0)$ harmonic superspace.

\section{Superform formulation of the covariant chiral anomaly}
\label{Sec4}

In our recent paper \cite{KNS}, we presented a superform
formulation of the anomalous current multiplet for the abelian
chiral anomaly. In this section, we generalise those results to
the case of the non-abelian chiral anomaly. In
particular, a superform formulation of the covariant anomalous
current multiplet will be developed. For the reader's convenience,
we start this section with a short review of the superform
descriptions of the 6D Yang-Mills and linear multiplets.


\subsection{The Yang-Mills multiplet}

In this subsection we review the superspace formulation for the 6D
$\cN=(1,0)$ Yang-Mills supermultiplet.
To describe a non-abelian vector multiplet, the covariant
derivative of Minkowski superspace $D_A = (\partial_a , D_\a^i)$
has to be replaced with a gauge-covariant one, \bea \cD_A := D_A +
\ri V_A ~. \label{SYM-derivatives} \eea Here the  gauge connection
one-form $V = \rd z^A V_A$ takes its values in the Lie algebra of
the Yang-Mills gauge group. The covariant derivative algebra is
\bea
[{\cD}_A, {\cD}_B \} &=&
 T_{AB}{}^C{\cD}_C
    + \ri F_{AB} \ ,
\label{gauged-algebra}
\eea
where the only non-vanishing torsion
is
\be
T_\a^i{}_\b^j{}^c = - 2 \, \ri \, \eps^{ij} (\g^c)_{\a\b}
\ee
and $F_{AB}$ are the components of the gauge covariant field
strength two-form $F = \rd z^B \wedge \rd z^A F_{AB}$. The
covariant derivatives and field strength may be written in a
coordinate-free way as follows
\be
\cD = \rd + \ri V \ , \quad F =
\rd V - \ri V \wedge V \ ,
\label{CovDer}
\ee
where we have introduced $\cD := \rd z^{A} \cD_{A}$. The field strength $F$
satisfies the Bianchi identity
\be
\cD F = \rd F + \ri V \wedge F
- \ri F \wedge V = 0 \quad \Longleftrightarrow \quad \cD_{[A} F_{B
C \} } - T_{[A B}{}^{D} F_{|D|C \} } = 0 \ .
\label{FBI}
\ee

The Yang-Mills gauge transformation acts on the gauge covariant
derivatives $\cD_A$ and a matter  superfield $U$ (transforming in
some representation of the gauge group) as
\be
\cD_A ~\rightarrow~
\re^{\ri  \t} \cD_A \re^{- \ri \t } , \qquad  U~\rightarrow~ U' =
\re^{\ri  \t} U~, \qquad \t^\dag = \t \ ,
\label{2.2}
\ee
where the Hermitian gauge parameter ${\t} (z)$ takes its values in the
Lie algebra of the gauge group. This implies that the gauge
connection and field strength transform as follows
\be
V \rightarrow  \re^{\ri \t } \,V \,\re^{- \ri \t} - \ri \re^{\ri \t
} \,\rd \re^{- \ri \t } \ , \quad F \rightarrow \re^{\ri \t } \,F
\,\re^{- \ri \t} \ .
\ee

Some components of the field strength two-form have to be
constrained in order to describe an irreducible multiplet. Upon
constraining the lowest mass dimension component of the field
strength two-form as
\bea
F_\a^i{}_\b^j = 0 \ , \quad F_a{}_\b^j =
(\g_a)_{\b \g} W^{j \g} \ ,
\eea
the remaining component is
determined from the Bianchi identity \eqref{FBI} to be
\bea F_{ab} = - \frac{\ri}{8} (\g_{ab})_\b{}^\a \cD_\a^k W^\b_k \ ,
\eea
and the superfield $W^{i\a }$ is required to obey the differential
constraints
\bea
\cD_\g^k W^\g_k = 0 \ , \quad \cD_\a^{(i} W^{j)\b
} = \frac{1}{4} \d_\a^\b \cD_\g^{(i} W^{j)\g }
\label{vector-Bianchies} \ .
\eea


\subsection{The superform formulation for the linear multiplet}

It is instructive to first describe the conserved current
multiplet in the abelian case which invariably is described by a
linear multiplet (or $\cO(2)$ multiplet). The linear multiplet can
be described using a four-form gauge potential $B = \frac{1}{4!}
\rd z^D \wedge \rd z^C \wedge \rd z^B \wedge \rd z^A B_{ABCD}$
possessing the gauge transformation \be \d B = \rd \r \ , \ee
where the gauge parameter $\r$ is an arbitrary  three-form. The
corresponding field strength is \be H = \rd B = \frac{1}{5!} \rd
z^E \wedge \rd z^{D} \wedge \rd z^C \wedge \rd z^B \wedge \rd z^A
H_{A B C D E} \ , \ee where \be H_{A B C D E} = 5 D_{[A}
B_{BCDE\}} - 10 T_{[A B}{}^{F} B_{|F|CDE\}} \ . \ee The field
strength must satisfy the Bianchi identity \be \rd H = 0  \quad
\Longleftrightarrow \quad D_{[A} H_{BCDEF \}} - \frac{5}{2}
T_{[AB}{}^{G} H_{|G|CDEF\}} = 0 \ . \ee

In order to describe the linear multiplet we need to impose some
covariant constraints on the field strength $H$. We choose the
constraint
\begin{align}
H_{abc}{}_\a^i{}_\b^j = - 2 \ri (\g_{abc})_{\a\b} L^{ij} \ , \quad
L^{ij} = L^{ji} \ , \label{O(2)constrants}
\end{align}
and require all lower dimension components to vanish. We can now
solve for the remaining components of $H$ in terms of $L^{ij}$.
The solution is \bea H_{abcd}{}_\a^i &=& - \frac{1}{6}
\eps_{abcdef} (\g^{ef})_\a{}^\b D_{\b j} L^{ij}
\ , \\
H_{abcde} &=& - \frac{\ri}{24} \eps_{abcdef}
({\tilde{\g}}^f)^{\a\b} D_\a^k D_\b^l L_{kl} \ , \eea where
$L^{ij}$ is required to satisfy the constraint for the linear
multiplet \be D_{\a}^{(i} L^{jk)} = 0 \ . \ee


\subsection{The superform formulation for the non-abelian current multiplet}

A non-abelian current multiplet described by a superfield $L^{ij}
= L^{(ij)}$ must take values in the Lie algebra of the gauge group
and satisfy the constraint \be \cD_\a^{(i} L^{jk)} = 0 \ .
\label{covConstLij} \ee To find a superform formulation we need to
introduce a five-form $\cH$ built out of $L^{ij}$ such that its
superform equation is satisfied as a result of the conservation
equation above. To do this we write down the superform equation
\be \cD \cH - \S = 0 \ , \label{nonAcSuperform1} \ee where $\S$ is
some covariant six-form taking values in the Lie algebra of the
gauge group. Consistency of the above equation requires \be \cD \S
- [F , \cH] = 0 \implies \cD_{[A_1} \S_{A_2 \cdots A_7 \}} - 3
T_{[A_1 A_2}{}^{B} \S_{|B|A_3 \cdots A_7 \} } = 3 [F_{[A_1 A_2} ,
\cH_{A_3 \cdots A_7 \} }]  \ .  \label{nonAcSuperform2} \ee The
equation \eqref{nonAcSuperform2} does not have a bosonic analogue
since a seven-form in six dimensions vanishes, but it becomes an
important requirement in the supersymmetric case where it demands
that there exists a covariant solution to equation
\eqref{nonAcSuperform2}.\footnote{This requirement is known in the literature as
Weil triviality \cite{BPT87}.} One can check
that there exists a covariant solution to \eqref{nonAcSuperform2}
and its non-vanishing components are: \bsubeq \bea
\S_{abcde}{}_\a^i &=& - \eps_{abcdef} (\g^f)_{\a\b} [W^\b_j ,
L^{ij} ]
\ , \\
\S_{abcdef} &=& \frac{\ri}{16} \eps_{abcdef} \Big( [\cD_\a^{i}
W^{j \a} , L_{ij}]  - \frac{8}{3} \{ W^{\a}_i , \cD_{\a j} L^{ij}
\} \Big) \ .
\eea
\esubeq
The superform $\cH$ possesses the
following non-vanishing components:
\bsubeq
\bea
\cH_{abc}{}_\a^i{}_\b^j
&=& - 2 \ri (\g_{abc})_{\a\b} L^{ij} \ , \\
\cH_{abcd}{}_\a^i &=& - \frac{1}{6} \eps_{abcdef}
(\g^{ef})_\a{}^\b \cD_{\b j} L^{ij}
\ , \\
\cH_{abcde} &=& - \frac{\ri}{24} \eps_{abcdef}
({\tilde{\g}}^f)^{\a\b} \cD_\a^k \cD_\b^l L_{kl} \ .
\eea
\esubeq
The superform equations \eqref{nonAcSuperform1} and
\eqref{nonAcSuperform2} are satisfied as a consequence of the
constraint \eqref{covConstLij}.


\subsection{The superform formulation for the anomalous non-abelian current multiplet}

To describe the covariant anomaly one only needs to modify the
superform equation \eqref{nonAcSuperform1} as follows
\be
\cD \cH - \S = \kappa \, {\rm str} (T^\cA F \wedge F \wedge F) T_{\cA} =
\kappa \, d^{\cA\cB\cC\cD} F_{\cA} \wedge F_{\cB} \wedge F_\cC \,
T_{\cD} \ , \label{defomedSFEq}
\ee
where $\kappa$ is some
constant and `${\rm str}$' is the symmetrised trace.
The two-form field strength $F$ takes its values in the Lie algebra of the gauge group
with generators $T^{\cA}$
\be
F = F_{\cal A} T^{\cA} \ ,
\ee
and $d^{\cA\cB\cC\cD}$ is the gauge-invariant tensor
\be
d^{\cal ABCD} = \tr(T^{(\cal A} T^{\cal B} T^{\cal C} T^{{\cal D})}) ~.
\label{dTensorInv}
\ee

The solution to the superform equation \eqref{defomedSFEq} is just
a deformation of the solution in the previous subsection. It turns
out that only two components of $\cH$ must be modified and they are
given by \bsubeq \bea \cH_{abcd}{}_\a^i &=&
- \frac{1}{6} \eps_{abcdef} (\g^{ef})_\a{}^\b \cD_{\b j} L^{ij} \non \\
&&+ \, \kappa \, \ri\, \eps_{abcdef} (\g^e)_{\a\b} (\g^f)_{\g\d}
\, {\rm str} \Big(T^\cA  W^\b_j W^{(j \g } W^{i) \d } \Big) T_\cA
\ , ~~~~~\\
\cH_{abcde} &=& \eps_{abcdef} \tilde{\cH}^f \ , \eea \esubeq where
\bea \tilde{\cH}^a &=& - \frac{\ri}{24} ({\tilde{\g}}^a)^{\a\b}
\cD_\a^k \cD_\b^l L_{kl}
+ \frac{\kappa}{8} {\rm str} \Big(T^\cA (\cD_{\g k} W^\g_{l}) (W^{k} \g^a W^{l}) \Big)T_\cA \non\\
&&\qquad+ \frac{3 \kappa \, \ri}{8} {\rm str} \Big( T^\cA F_{bc}
(W^{ k} \g^{abc} W_k) \Big) T_\cA \ .~~~ \eea The superform
equation \eqref{defomedSFEq} requires that $L^{ij}$ satisfy the
differential equation
\bea
\cD_\a^{(i} L^{jk)} &=& \kappa \, \ri
\, \eps_{\a\b\g\d} \, {\rm str} \Big( W^{(i \b} W^{j \g} W^{k) \d}
T^\cA
\Big) T_\cA \non\\
&=& \kappa \, \ri \, \eps_{\a\b\g\d} \, d^{\cA\cB\cC\cD} W^{(i
\b}_{\cA} W^{j \g}_{\cB} W^{k) \d}_{\cC} \, T_{\cD} \ .
\label{2.28}
\eea
The value of the constant $\kappa$ will be fixed by the explicit calculation in the
next section. In particular, for the model of a hypermultiplet
interacting with a background Yang-Mills multiplet this constant is
$\kappa = -\frac{1}{96 \pi^3}$.

The highest dimension component of \eqref{defomedSFEq} implies
the following
\be
\cD_a \tilde{\cH}^a - \frac{\ri}{16}  \Big(
[\cD_\a^{i} W^{j \a} , L_{ij}]  - \frac{8}{3} \{ W^{\a}_i ,
\cD_{\a j} L^{ij} \} \Big) = \frac{\kappa}{8} \eps^{abcdef} {\rm
str} (T^{\cA} F_{ab} F_{cd} F_{ef}) T_{\cA}  \ .
\ee
Thus the
component projection of $\tilde{\cH}^a$ should be understood as
the current whose conservation condition is now deformed.


\subsection{Another deformation of the non-abelian current multiplet}

It is worth mentioning that besides the deformation just
considered, there exists another deformation of the current
multiplet. To describe it, one modifies the superform equation
\eqref{nonAcSuperform1} as follows
\be
\cD \cH - \S = \mu \, \tr
(F \wedge F) \wedge F \ ,
\label{defomed2SFeq}
\ee
where $\mu$ is
some constant. The superform equation \eqref{defomed2SFeq} is
solved by
\bsubeq
\bea \cH_{abc}{}_\a^i{}_\b^j
&=& - 2 \ri (\g_{abc})_{\a\b} L^{ij} \ , \\
\cH_{abcd}{}_\a^i &=& - \frac{1}{6} \eps_{abcdef}
(\g^{ef})_\a{}^\b \cD_{\b j} L^{ij} + \mu \, \ri \, \eps_{abcdef}
(\g^e)_{\a\b} (\g^f)_{\g\d} \tr(W^{(i \g} W^{j) \d}) W^\b_j
\ , ~~~~~\\
\cH_{abcde} &=& \eps_{abcdef} \tilde{\cH}^f \ ,
\eea
\esubeq
where
\be
\tilde{\cH}^a = - \frac{\ri}{24} ({\tilde{\g}}^a)^{\a\b}
\cD_\a^k \cD_\b^l \cL_{kl} + \frac{\mu}{8} \tr(W^k \g^a W^l)
\cD_{\g k} W^{\g}_l + \frac{3 \mu\, \ri}{8} \tr(W^k \g^{abc} W_k)
F_{bc}
\ee
and all lower dimension components of $\cH$ vanish. The
superfield $L^{ij}$ is now required to satisfy the
differential equation
\be
\cD_\a^{(i} L^{jk)} = \mu \, \ri \,
\eps_{\a\b\g\d} \tr(W^{(i \b} W^{j \g}) W^{k) \d} \ .
\label{2.29}
\ee
The highest dimension component of eq. \eqref{defomed2SFeq} implies
\be
\cD_a \tilde{\cH}^a - \frac{\ri}{16}  \Big( [\cD_\a^i W^{j \a}
, L_{ij}]  - \frac{8}{3} \{ W^{\a}_i , \cD_{\a j} L^{ij} \} \Big)
= \frac{\mu}{8} \eps^{abcdef} \tr(F_{ab} F_{cd}) F_{ef}  \ .
\ee
The component projection of $\tilde{\cH}^a$ corresponds to the
current with a deformed conservation equation as described above.


\section{Chiral anomaly in $\cN=(1,0)$ harmonic superspace}
\label{SecChirAnom}

In this section we discuss the general aspects of covariant and consistent anomalies
in 6D $\cN=(1,0)$ supersymmetric gauge theories.

\subsection{Effective action in non-anomalous gauge theories}

Let us consider an anomaly-free gauge theory in $\cN=(1,0)$ harmonic
superspace. The effective action $\Gamma$ of such a theory may
always be chosen to be
a gauge-invariant functional, $\G = \G [V^{++}]$,
 of the analytic gauge prepotential
$V^{++} = V^{++}_\cA T^\cA$
taking its values in the Lie algebra of the gauge group.
The generators of the gauge group, $T^{\cal A}$,
will be  normalised
so that $\tr_{(\rm F)}(T^\cA T^\cB) = \d^{\cA \cB}$ in the fundamental representation.
If the prepotential is perturbed, $V^{++} \to V^{++} + \d V^{++}$,
the variation of the effective action can be represented in the form
\begin{subequations}
\bea
\delta\Gamma &=&
 \int {\rm d}\zeta^{(-4)}\, \delta V^{++}_\cA L^{++ \cA}
 = \tr_{(\rm F)}  \int {\rm d}\zeta^{(-4)}\, \delta V^{++} L^{++ }
 ~,
\label{L-a}
\\
{D}^+_\alpha L^{++} &=& 0~,
\label{L-analyt}
\eea
\end{subequations}
for some effective  current $L^{++} = L^{++}_\cA T^\cA =L^{++}(V^{++})$.
The effective action is  invariant under infinitesimal gauge transformations
\be
\delta_\l V^{++} = -{\cal D}^{++} \lambda~,
\label{deltaV}
\ee
where the gauge parameter $\l =\l_\cA T^\cA $ is analytic, $D^+_\a \l=0$.

The invariance condition  $\d_\l \G=0$
implies that the effective current obeys the harmonic shortness constraint
\be
{\cal D}^{++} L^{++} = 0~.
\label{nabla-L}
\ee
The general solution to this constraint reads
\bea
L^{++} = {\rm e}^{\ri b} L^{++}_\tau {\rm e}^{-\ri b}~,\qquad
L^{++}_\tau(z,u) = L^{ij}(z) u^+_i u^+_j~,
\label{Luu}
\eea
where $ L^{ij}(z) $ obeys, as a consequence of \eqref{L-analyt},
 the conservation equation
\bea
\cD^{(i}_\alpha L^{jk)} =0~.
\eea
Here $b=b(z,u)$ is the bridge superfield, see Appendix A for the technical details.

The analyticity constraint ${D}^+_\alpha  V^{++}=0$ can always be solved as
\be
 V^{++} = ({ D}^+)^4  M^{--}~,
\label{V-M}
\ee
where $ M^{--}$ is an unconstrained superfield (subject to a certain reality condition)
on the full harmonic superspace.
It is defined modulo gauge transformations
\be
\delta_\xi M^{--} = {D}^+_\alpha \xi^{(-3)\alpha}~,
\label{delta-xi-M}
\ee
which do not change $V^{++}$ for
any unconstrained gauge parameter $\xi^{(-3)\alpha}$.
We emphasise that \eqref{deltaV} and \eqref{delta-xi-M}
are two different gauge symmetries.
The gauge transformation  \eqref{delta-xi-M} is absent when one works solely with
$V^{++}$.

The vector multiplet can be described either in terms of
$V^{++}$ or in terms of $M^{--}$. We will refer to these descriptions as
$V$-formulation and $M$-formulation, respectively.
The gauge freedom in the $V$-formulation is given by \eqref{deltaV}.
Let us now discuss, in some more detail, the gauge freedom in the $M$-formulation.

When dealing with $M^{--}$, it is natural to express
the analytic gauge parameter $\lambda$  in \eqref{deltaV}
via an unconstrained one to be denoted $\rho^{(-4)}$,
\be
\lambda = ({D}^+)^4 \rho^{(-4)}~.
\ee
Then the $\l$-transformation (\ref{deltaV}) results in the following variation of $M^{--}$
\be
\delta_\lambda M^{--} = - {\cal D}^{++} \rho^{(-4)}~,
\label{delta-lambda-M}
\ee
modulo a $\x$-transformation \eqref{delta-xi-M}.
The complete gauge freedom in the $M$-formulation is
\bea
\d M^{--} = -  {\cal D}^{++} \rho^{(-4)}
+{D}^+_\alpha \xi^{(-3)\alpha}~.
\label{M-gauge-freedom}
\eea

The prepotential $M^{--}$ has the following harmonic expansion:
\bea
M^{--}(z,u) &=& M^{ij}(z) u^{-}_i u^{-}_j
+\sum_{\k=1}^{\infty} M^{(i_1 \dots i_k j_1 \dots j_{k+2})}(z)
u^+_{i_1} \dots u^+_{i_k} u^-_{j_1} \dots u^-_{j_{k+2}} ~.
\label{2.10}
\eea
A similar series can be introduced for the gauge parameter $\rho^{(-4)}$
\bea
\r^{(-4)}(z,u) &=&
\sum_{\k=0}^{\infty} \r^{(i_1 \dots i_k j_1 \dots j_{k+4})}(z)
u^+_{i_1} \dots u^+_{i_k} u^-_{j_1} \dots u^-_{j_{k+4}} ~.
\label{2.12}
\eea
Then the transformation law \eqref{delta-lambda-M} tells us that
all the superfields $M^{(i_1 \dots i_k j_1 \dots j_{k+2})}$, $k \geq 1$,
in the Fourier series \eqref{2.10} can be gauged away algebraically. In the resulting gauge
\bea
M^{--}(z,u) = M^{ij}(z) u^{-}_i u^{-}_j~,
\label{2.11}
\eea
the local symmetry \eqref{delta-lambda-M} is completely fixed.
However, we still have the freedom to perform  $\x$-transformations
  \eqref{delta-xi-M} generated by a single harmonic-independent
parameter $\xi^{ijk\,\alpha}(z)$, which originates as
$\xi^{(-3)\alpha}(z,u) =
\frac43\xi^{ijk\,\alpha}(z) u^-_i u^-_j u^-_k$.
The gauge superfield $M^{ij} (z)$ may be recognised as
Mezincescu's prepotential  \cite{HST,Koller,Mezincescu}.
We will refer to the  unconstrained superfield $M^{--}(z,u)$ defined by
\eqref{V-M} as the generalised Mezincescu prepotential.

The above discussion shows that
all the superfields $M^{(i_1 \dots i_k j_1 \dots j_{k+2})}$, $k \geq 1$,
in the Fourier series \eqref{2.10} may be interpreted as compensators, for
all of them  can be  gauged away algebraically by applying
local transformations  \eqref{delta-lambda-M}.\footnote{One of the most familiar examples of compensators is the scalar field of the Stueckelberg formulation,
which is used to introduce a local gauge invariance in the theory with
a massive vector field.}
Thus in the $M$-formulation, the local $\xi$-invariance (\ref{delta-xi-M})
plays the role of genuine gauge transformations
while the $\r$-gauge freedom (\ref{delta-lambda-M}) becomes purely compensating.
The significance of this observation is that, in general, the compensating gauge
symmetries are known to be non-anomalous \cite{deWG}.
This means that in the $M$-formulation the presence of chiral anomalies
is equivalent to the fact that  the local $\xi$-invariance (\ref{delta-xi-M})
becomes anomalous.

For our subsequent discussion,
it is instructive to look at the gauge transformations of $M^{--}$ in  the $\t$-frame.
By construction, the relation (\ref{V-M}) is defined in the $\lambda$-frame, where the
gauge-covariant spinor derivative ${\cal D}^+_\alpha$ has no gauge connection,
${\cal D}^+_\alpha=D^+_\alpha$.
If the analytic prepotential is
subject to a perturbation,
$V^{++} \to V^{++} + \d V^{++}$, then the generalised Mezincescu prepotential
also changes, $M^{--} \to M^{--} + \d M^{--}$, such that
\be
\delta V^{++} = ({D}^+)^4 \delta  M^{--}~.
\label{delta-V-M}
\ee
This relation in the $\tau$-frame becomes
\be
\delta V^{++}_\tau = ({\cal D}^+)^4 \delta  M^{--}_\tau~,
\ee
where
\be
\delta V^{++}_\tau = {\rm e}^{-\ri b} \delta V^{++} {\rm e}^{\ri b}~,\qquad
\delta M^{--}_\tau = {\rm e}^{-\ri b} \delta M^{--} {\rm e}^{\ri b}~.
\ee
For $\delta M^{--}_\tau$ the gauge transformations (\ref{delta-xi-M}) and (\ref{delta-lambda-M}) read, respectively,
\begin{subequations}
\bea
\delta_\xi M^{--}_\tau &=& {\cal D}^+_\alpha \xi^{(-3)\alpha}_{\tau}~,\label{a}\\
\delta_\lambda M^{--}_\tau &=& -D^{++} \rho^{(-4)}_\tau~,
\label{b}
\eea
\end{subequations}
where
\be
\xi^{(-3)\alpha}_{\tau} = {\rm e}^{-\ri b}
 \xi^{(-3)\alpha} {\rm e}^{\ri b}~,\qquad
\rho^{(-4)}_\tau = {\rm e}^{-\ri b} \rho^{(-4)}
 {\rm e}^{\ri b}~.
\ee
Unlike  the original gauge transformations (\ref{delta-xi-M}) and (\ref{delta-lambda-M})
in the $\l$-frame, the spinor derivative in \eqref{a} is gauge covariant,
while the harmonic derivative in \eqref{b} has no gauge connection.

In the $\tau$-frame, the superfields $\delta M^{--}_\tau$, $\xi^{(-3)\alpha}_\tau$
and $\rho^{(-4)}_\tau$
have the following harmonic expansions:
\bea
\delta M^{--}_\tau(z,u) &=&
\delta M^{ij}(z) u^{-}_i u^{-}_j
+\sum_{\k=1}^{\infty} \d M^{(i_1 \dots i_k j_1 \dots j_{k+2})}(z)
u^+_{i_1} \dots u^+_{i_k} u^-_{j_1} \dots u^-_{j_{k+2}}
~,\\
\xi^{(-3)\alpha}_\tau(z,u) &=&
\frac43\xi^{ijk\,\alpha}(z) u^-_i u^-_j u^-_k +
\ldots~,\\
\rho^{(-4)}_\tau(z,u)&=&
\sum_{\k=0}^{\infty} \r_\t^{(i_1 \dots i_k j_1 \dots j_{k+4})}(z)
u^+_{i_1} \dots u^+_{i_k} u^-_{j_1} \dots u^-_{j_{k+4}} ~.
\label{rho-series}
\eea
It is clear that the gauge freedom \eqref{b} may be used to impose a gauge condition
\bea
\delta M^{--}_\tau(z,u) &=&
\delta M^{ij}(z) u^{-}_i u^{-}_j~.
\label{2.22}
\eea
The residual gauge transformations,  which preserve the gauge,
are generated by
\bea
\xi_\t^{(-3)\alpha}(z,u) =
\frac43\xi^{ijk\,\alpha}(z) u^-_i u^-_j u^-_k ~, \qquad
\r^{(-4)} (z,u) = \frac{1}{3} {\cal D}^i_\alpha \xi^{jkl \a}u^-_i u^-_j u^-_k u^-_l~.
\eea
In accordance with \eqref{a}, the Mezincescu prepotential transforms as
\be
\delta_\xi M_{ij} = {\cal D}^k_\alpha \xi^\alpha_{ijk}~.
\label{xi-M}
\ee

One can use the operator
$({D}^+)^4$ in (\ref{delta-V-M}) to restore the full superspace measure
in (\ref{L-a}) and to represent the variation of the effective
action in two equivalent forms
\bea
\delta \Gamma =
\tr_{(\rm F)} \int {\rm d}^{6|8}z {\rm d}u\, \delta M^{--} L^{++ }
= \tr_{(\rm F)} \int {\rm d}^{6|8}z {\rm d}u\, \delta M^{--}_\t L^{++}_\t
~.
\label{deltaM-Gamma}
\eea
where the effective current $L^{++}_\t$ is given by eq.\ (\ref{Luu}).
In the $\t$-frame, the
harmonic integral can be easily computed to result with
\be
\delta \Gamma = \frac13
\int {\rm d}^{6|8}z\, \delta M^{ij}_\cA L_{ij}^\cA~.
\label{ML}
\ee
The invariance of the effective action under
the $\xi$-transformations (\ref{xi-M}), $\delta_\xi\Gamma=0$, is equivalent to
the analyticity constraint on the effective current $L^{ij}$,
\be
{\cal D}^{(i}_\alpha L^{jk)} =0~.
\label{Lanal}
\ee
The $\lambda$ (or $\r$) gauge freedom is completely gone once the gauge
condition \eqref{2.22} has been chosen.


\subsection{Chiral anomaly and deformed conservation laws}

As was pointed out in the previous subsection, in non-anomalous gauge
theories the effective current $L^{++}$ obeys the constraints of
Grassmann analyticity (\ref{L-analyt}) and harmonic shortness
(\ref{nabla-L}). Either of these constraints may, in principle, be
violated in theories which suffer from chiral anomalies depending on
which of the gauge transformations, (\ref{deltaV}) or
(\ref{delta-xi-M}), becomes broken. We recall that eq.\ (\ref{deltaV})
describes the gauge freedom in the $V$-formulation
while in the $M$-formulation the gauge freedom is larger
and is given by \eqref{M-gauge-freedom}.
In the latter case the $\r$-transformation is compensating
and, therefore, non-anomalous \cite{deWG}.
It is the $\x$-transformations which are anomalous in the $M$-formulation.

In those supersymmetric gauge theories which suffer from chiral
anomalies at the quantum level, the $V$-formulation and the $M$-formualtion
become non-equivalent as they are described by two different effective
currents which we denote by $L^{++}_{\rm an}$ and $L^{++}_{\rm Mez}$,
respectively.

In the $V$-formulation, the effective
current $L^{++}_{\rm an}$ remains analytic while the harmonic
shortness constraint (\ref{nabla-L}) may be broken
\be
{\cal D}^+_\alpha L^{++}_{\rm an} =0
~,\qquad
{\cal D}^{++} L^{++}_{\rm an} = A^{(+4)}~.
\label{A+4}
\ee
In contrast, in the $M$-formulation
the effective current $L^{++}_{\rm Mez}$
must obey the harmonic shortness condition
(\ref{nabla-L}) while the Grassmann analyticity constraint may be
deformed
\be
{\cal D}^{++} L^{++}_{\rm Mez} = 0~,\qquad
{\cal D}^{+}_\alpha L^{++}_{\rm Mez} = A_\alpha^{(+3)}~.
\label{L-non-anal}
\ee
Here $A^{(+4)}$ and $A_\alpha^{(+3)}$ are some composite operators of the vector multiplet which,
in the non-abelian case, must obey the Wess-Zumino consistency
condition \cite{WZ-cons} (see the next subsection).

Let us denote the difference between $L^{++}_{\rm an}$ and $L^{++}_{\rm
Mez}$ by $\tilde L^{++}$,
\be
\tilde L^{++} = L^{++}_{\rm an} - L^{++}_{\rm Mez}~.
\label{tildeL}
\ee
By construction, this superfield obeys
\be
{\cal D}^{++} \tilde L^{++} = A^{(+4)}~,\qquad
{\cal D}^{+}_\alpha \tilde L^{++} = -A_\alpha^{(+3)}~.
\ee
Thus, given the superfield $\tilde L^{++}$, one could transfer the
chiral anomaly from one formulation to the other.

According to the results of the previous section, see eq.\ (\ref{2.28}),
the admissible deformation of analyticity of the effective current
reads
\be
A^{(+3)}_\alpha = \ri\,\kappa\, \varepsilon_{\alpha\beta\gamma\delta}
W^{+\beta} W^{+\gamma} W^{+\delta} ~.
\label{A3-KNS}
\ee
One of the aims of this work is to derive this expression for the
chiral anomaly by analysing the short distance behaviour
of the hypermultiplet propagator.
The explicit form of $A^{(+4)}$
will be be discussed in section \ref{Summary}.

\subsection{Consistent chiral anomaly in harmonic superspace}
\label{Sec2.3}

In this subsection we make use of the $M$-formulation in which
the variation of the effective action is given by the full-superspace
integral (\ref{deltaM-Gamma}), and the chiral anomaly
appears as a deformation of the Grassmann analyticity constraint
for the effective current (\ref{L-non-anal}). Since the $V$-formulation
will not  be discussed in this subsection, we omit the
subscript `Mez' assuming that we always
work with the $M$-formulation, $L^{++}_{\rm Mez}\equiv L^{++}$.

Let us consider the variation of the effective action
(\ref{deltaM-Gamma}), where $\delta M^{--}$ is the $\xi$-gauge
transformation (\ref{delta-xi-M}). The Wess-Zumino consistency
condition \cite{WZ-cons} for this variation implies
\be
(\delta_{\xi_1}\delta_{\x_2}-\delta_{\x_2}\delta_{\x_1})\Gamma =
\delta_{[\xi_1,\x_2]}\Gamma~,
\label{consist}
\ee
where $\xi_1$ and $\x_2$ are two gauge parameters taking values in the Lie algebra of the gauge group.
In the non-abelian case the Wess-Zumino consistency
condition (\ref{consist}) becomes a non-trivial constraint for the
effective action which may, in principle, be solved using the
descent equation approach \cite{ZuminoZee}. In this section, however,
to construct the consistent chiral anomaly we will follow the ideas
of Leutwyler \cite{Leutwyler} generalised to the superfield
formalism. Note that Leutwyler's approach has proved to be  very efficient for
obtaining the consistent anomalies of 4D $\cN=1$ supersymmetric gauge theories in
superspace \cite{OMc,J}.

Let $T^{\cal A}$ be the generators of the gauge group. The prepotential
$M^{--}$ may be written as a linear combination of the generators,
\be
M^{--}(z,u) = M^{--}_{\cal A}(z,u) T^{\cal A}~.
\ee
Then, the variation (\ref{deltaM-Gamma}) can be cast in the form
\be
\delta\Gamma[M^{--}] = \int {\rm d}^{6|8} z{\rm d}u \, \delta
M^{--}_{\cal A}
L^{++{\cal A}}(M^{--})~,
\label{Gamma-var}
\ee
where
\be
L^{++\cal A} = \tr_{\rm (F)}( T^{\cal A} L^{++})~.
\ee
Here the effective current $L^{++}$ is treated as a function of the
prepotential $M^{--}$ (possibly, with superspace derivatives). The
variation (\ref{Gamma-var}) is integrable provided that
the effective current obeys
\be
\frac{\delta L^{++\cal B}(z_2,u_2)}{\delta M_{\cal A}^{--}(z_1,u_1)}
-\frac{\delta
L^{++\cal A}(z_1,u_1)}{\delta M_{\cal B}^{--}(z_2,u_2)} = 0~.
\label{cons--cond}
\ee
For such an effective current the variation (\ref{Gamma-var}) may be
formally integrated
\be
\Gamma[M^{--}] = \int {\rm d}^{6|8} z{\rm d}u \, M^{--}_{\cal A}
\int_0^1 {\rm d}y\,
L^{++\cal A}(yM^{--})~.
\label{Gamma-y}
\ee

For anomalous gauge theories, however, direct quantum computations
usually result in an effective current which fails to satisfy
(\ref{cons--cond}) in the non-abelian case.
We denote such a current by $L^{++}_{\rm
cov}$, emphasising that ${\cal D}^+_\alpha L^{++}_{\rm cov}=A^{(+3)}_\alpha$
is a gauge-covariant superfield.
The corresponding part of the variation of effective action with
this current reads
\be
\delta\Gamma_{\rm cov}[M^{--}] = \int {\rm d}^{6|8} z{\rm d}u \, \delta
M^{--}_{\cal A}
L_{\rm cov}^{++\cal A}(M^{--})~.
\label{Gamma-inconsistent}
\ee
Given the effective current $L^{++}_{\rm cov}$, one can still
construct a functional $\Gamma[M^{--}]$ by the rule
(\ref{Gamma-y}),
\be
\Gamma[M^{--}] = \int {\rm d}^{6|8} z{\rm d}u \, M^{--}_{\cal A}
\int_0^1 {\rm d}y\,
L^{++\cal A}_{\rm cov}(yM^{--})~.
\label{Gamma-y1}
\ee
The general variation of this functional differs from
(\ref{Gamma-inconsistent}) by the consistency terms
which we denote here by $\delta\Gamma_{\rm cons}$,
\bea
\delta\Gamma[M^{--}] &=& \delta\Gamma_{\rm cov}[M^{--}] +
\delta\Gamma_{\rm cons}[M^{--}]~,
\label{Gamma-full}
\eea
where
\begin{subequations}
\label{dGamma-cov}
\bea
\delta\Gamma_{\rm cons}[M^{--}] &=&  \int_0^1 {\rm d}y\, X(y)~,
\label{X-terms-0}\\
X(y) &=& y\int {\rm d}^{6|8}z_1 {\rm d}u_1 {\rm d}^{6|8}z_2 {\rm d}u_2\,
\delta M^{--}_{\cal A}(z_1,u_1) M_{\cal B}^{--}(z_2,u_2)\non\\&&\times
\left[
\frac{\delta L^{++\cal B}_{\rm cov}(yM^{--}(z_2,u_2))}{\delta(yM_{\cal A}^{--}(z_1,u_1))}
-\frac{\delta L^{++\cal A}_{\rm cov}(yM^{--}(z_1,u_1))}{\delta(yM_{\cal B}^{--}(z_2,u_2))}
\right]\,.
\label{X-terms}
\eea
\end{subequations}
Obviously, the variation (\ref{Gamma-full}) is integrable as it is
derived from the functional (\ref{Gamma-y1}). Thus, given an
effective current $L^{++}_{\rm cov}$ which does not satisfy the
consistency condition (\ref{cons--cond}), it is always possible to
construct the consistency terms (\ref{dGamma-cov}) such that the
variation (\ref{Gamma-full}) becomes integrable.

The relation between the analytic and Mezincescu's prepotentials
(\ref{V-M}) may be used to prove the useful identity
\be
\frac{\delta L^{++\cal B}(z_2,u_2)}{\delta M^{--}_{\cal A}(z_1,u_1)} = \frac{\delta
L^{++\cal B}(z_2,u_2)}{\delta V^{++}_{\cal A}(z_1,u_1)}~.
\ee
This identity allows us to represent the consistency terms (\ref{X-terms})
in the equivalent form
\bea
X(y) &=& y\int {\rm d}^{6|8}z_1 {\rm d}u_1 {\rm d}^{6|8}z_2 {\rm d}u_2\,
\delta M^{--}_{\cal A}(z_1,u_1) M_{\cal B}^{--}(z_2,u_2)\non\\&&\times
\left[
\frac{\delta L^{++\cal B}_{\rm cov}(y V^{++}(z_2,u_2))}{\delta(y V_{\cal A}^{++}(z_1,u_1))}
-\frac{\delta L^{++\cal A}_{\rm cov}(y V^{++}(z_1,u_1))}{\delta(y V_{\cal B}^{++}(z_2,u_2))}
\right]\,.
\label{X-term1}
\eea
Here we consider the effective current as a function of the
analytic prepotential $V^{++}$ rather than  the generalised Mezincescu
prepotential $M^{--}$.

In this subsection, we have so far been considering
the general variation $\delta M^{--}$.
The consistent chiral anomaly appears when the $\xi$-gauge
variation (\ref{delta-xi-M}) is substituted in (\ref{Gamma-full}).
In particular, eq.\ (\ref{Gamma-inconsistent}) turns into
\bea
\delta_\xi\Gamma_{\rm cov}
&=& \int {\rm d}^{6|8} z{\rm d}u \,
\xi_{\cal A}^{(-3)\alpha} A^{(+3)\cal A}_\alpha~,
\label{Gamma-cov1}
\eea
where
\be
A^{(+3)} = {\cal D}^+_\alpha L_{\rm cov}^{++} ~,
\label{2.39}
\ee
and $\xi^{(-3)\alpha} = \xi^{(-3)\alpha}_{\cal A} T^{\cal A}$.
The consistency terms $\delta_\xi\Gamma_{\rm cons}$ have the form
(\ref{X-terms-0}) with
\be
X(y) =- y\int {\rm d}^{6|8}z_1 {\rm d}u_1 {\rm d}^{6|8}z_2 {\rm d}u_2\,
\xi^{(-3)\alpha}_{\cal A}(z_1,u_1)
M^{--}_{\cal B}(z_2,u_2)
\frac{\delta A^{(+3)\cal A}_\alpha (yV^{++}(z_1,u_1))}{\delta(yV^{++}_{\cal B}(z_2,u_2))}
~.
\label{X-term2}
\ee
The relation (\ref{X-term2}) follows from (\ref{X-term1}) upon
substituting the $\xi$-gauge variation (\ref{delta-xi-M}),
integrating by parts the derivative ${\cal D}^+_\alpha$ and taking
into account that only the last term in brackets in
(\ref{X-term1}) may contribute since the other term is analytic in
$(z_1,u_1)$ by construction. The equation (\ref{2.39}) has also been applied.

We emphasise  that in this section we discussed the
properties of chiral anomaly for {\it general} $\cN=(1,0) $
supersymmetric gauge theories in
harmonic superspace. In the next section we will explicitly
compute the covariant anomaly $A^{(+3)}_\alpha$ and the
corresponding consistency terms for the model of a hypermultiplet
interacting with an external vector multiplet.


\section{Chiral anomaly for the  hypermultiplet effective action}
\label{Sec3}

In this section we analyse
the short-distance behaviour of the propagator for the hypermultiplet coupled to a
background vector multiplet, and then apply the results of this analysis
to  compute the covariant and consistent chiral anomalies.

\subsection{Hypermultiplet effective action}
\label{Sec2}

In harmonic superspace, the hypermultiplet is described by an analytic superfield $q^+$, ${\cal D}^+_\alpha q^+ =0$,
and its tilde-conjugate $\tilde q^+$.
The classical action of the hypermultiplet interacting with the gauge
superfield $V^{++}$ has the standard form \cite{GIKOS,HSWest}
\be
S= - \int {\rm d}\zeta^{(-4)} \, \tilde q^+ {\cal D}^{++} q^+~,\qquad
{\cal D}^{++} = D^{++} + \ri\, V^{++}~.
\label{S}
\ee
This action is invariant under infinitesimal gauge transformations
(\ref{deltaV}) with hypermultiplets transforming as
\be
\delta q^+ = \ri\,\lambda\, q^+ ~, \qquad
\delta\tilde q^+ = -\ri\, \tilde q^+ \lambda~,
\ee
where $\lambda$ is analytic gauge parameter.
The hypermultiplet is assumed to transform in some representation
of the gauge group.

The effective action in the hypermultiplet model (\ref{S}) is given by
\be
\Gamma = \ri\, \Tr \ln {\cal D}^{++} = -\ri\, \Tr \ln G^{(1,1)}~,
\label{Gamma}
\ee
where the functional trace `$\Tr$' corresponds to the space of analytic superfields of $\sU(1)$ charge +1,
and $G^{(1,1)}$ is the hypermultiplet propagator obeying the equation
\be
{\cal D}^{++}_1 G^{(1,1)}(1|2) = \delta_{\rm A}^{(3,1)}(1|2){\mathbbm 1}~.
\label{G-eq}
\ee
Here ${\mathbbm 1}$ is the unit matrix,
 and $\delta_{\rm A}^{(3,1)}(1|2)$ is the analytic delta-function which is related to the full superspace delta-function $\delta^{6|8}(z_1-z_2)$ as
\be
\delta_{\rm A}^{(q,4-q)}(1|2) = ({\cal D}^+_1)^4 \delta^{6|8}(z_1-z_2)\delta^{(q-4,4-q)}(u_1,u_2)~,\qquad
({\cal D}^+)^4:= -\frac1{96} \varepsilon^{\alpha\beta\gamma\delta}
{\cal D}^+_\alpha {\cal D}^+_\beta {\cal D}^+_\gamma
{\cal D}^+_\delta~.
\label{an-delta}
\ee
The solution to (\ref{G-eq}) is derived in complete analogy with
the 4D $\cN=2$ and 5D $\cN=1$ cases \cite{BFM,K-5D}.
The solution  \cite{BP} is
\be
G^{(1,1)}(1|2) = \frac1{\stackrel{\frown}{\square}}({\cal D}^+_1)^4
({\cal D}^+_2)^4 \frac{\delta^{6|8}(z_1-z_2)}{(u^+_1 u^+_2)^3}{\mathbbm 1}~,
\label{GG}
\ee
where $\stackrel{\frown}{\square}$ is the gauge covariant d'Alembertian operator which maps the space of covariantly analytic superfields into itself,
\be
\stackrel{\frown}{\square} = \frac12 ({\cal D}^+)^4 {\cal D}^{--} {\cal D}^{--}~.
\label{def-box}
\ee
Acting on an analytic superfield $\Phi_{\rm A}$, ${\cal D}^+_\alpha\Phi_{\rm A}=0$, it can equivalently be written as \cite{BP}
\be
\stackrel{\frown}{\square} \Phi_{\rm A}=
\left({\cal D}^a {\cal D}_a - W^{+\alpha}{\cal D}^-_\alpha
 + \frac14 ({\cal D}^-_\alpha W^{+\alpha})
 -\frac14 ({\cal D}^+_\alpha W^{+\alpha}){\cal D}^{--}\right)\Phi_{\rm A}~.
\label{box-anal}
\ee
It is instructive to compare this expression for the 6D $\cN=(1,0)$
analytic d'Alembertian  $\stackrel{\frown}{\square} $
with its 4D $\cN=2$ and 5D $\cN=1$ cousins \cite{BFM,KL}.

The definition of the effective action (\ref{Gamma}) is purely
formal, since the operator
${\cal D}^{++}$ maps the space of covariantly analytic
superfields of $\sU(1)$ charge $+1$ into the space of
covariantly analytic superfields of $\sU(1)$ charge $+3$.
However, the variation of the effective action
\be
\delta\Gamma= -\Tr\,\Big\{ \delta V^{++} G^{(1,1)} \Big\}
\label{3.9}
\ee
makes sense. Explicitly, it can be written in the form
\begin{subequations}
\bea
\delta\Gamma &=&\tr_{({\rm F})} \int \rd\zeta^{(-4)} \delta V^{++} L^{++}_{\rm
an}~,\\
L^{++}_{\rm an} &=& -  G^{(1,1)}(1|2)|_{1=2}~.
\label{L-b}
\eea
\end{subequations}
The definition \eqref{3.9} is completely analogous to those in the 4D $\cN=2$
\cite{KM}
and 5D $\cN=1$ cases \cite{K-5D}. The specific feature of the 6D case
is that the variation \eqref{3.9} is not integrable due to chiral anomalies.

We point out that the effective current (\ref{L-b}) is analytic
owing to the analyticity of Green's function (\ref{GG}) with
respect to both arguments. Starting from this expression for the
propagator, it is possible to construct the effective current $L^{++}_{\rm Mez
}$ which is not analytic but satisfies the harmonic shortness
constraint (\ref{L-non-anal}). For this purpose we revisit the
form of the propagator in harmonic superspace developed in
\cite{KM}.

Recall that in the $\tau$-frame the covariant spinor derivative ${\cal D}^+_\alpha$ is linear in harmonic variables,
${\cal D}^+_\alpha = {\cal D}^i_\alpha u^+_i$. For this derivative one can
prove a simple identity
\be
{\cal D}^+_{2\,\alpha} \delta^{6|8}(z_1-z_2)
=[ -(u^+_1 u^+_2) {\cal D}^-_{1\,\alpha}
 +(u^-_1 u^+_2) {\cal D}^+_{1\,\alpha} ] \delta^{6|8}(z_1-z_2)~.
\label{D-delta-identity}
\ee
This identity, together with the algebra of gauge covariant derivatives
(\ref{algebra}), allows one to derive the useful property
\bea
({\cal D}^+_1)^4({\cal D}^+_2)^4 \frac{\delta^{6|8}(z_1-z_2)}{(u^+_1 u^+_2)^q}
{\mathbbm 1}&=&
({\cal D}^+_1)^4 \bigg[\frac{1}{(u^+_1 u^+_2)^{q-4}} ({\cal D}^-_1)^4
+
\frac{(u^-_1 u^+_2)}{(u^+_1u^+_2)^{q-3}}
 \Delta^{--}_1
 \label{idDD}
\\&+& \stackrel{\frown}{\square}_1\frac{(u^-_1 u^+_2)^2}{(u^+_1 u^+_2)^{q-2}}
+\frac{3-q}4\frac{(u^-_1 u^+_2)^3}{(u^+_1 u^+_2)^{q-1}}
({\cal D}^+_\alpha W^{+\alpha})
\bigg]
\delta^{6|8}(z_1 - z_2){\mathbbm 1}~,\non
\eea
where
\be
\Delta^{--} = -\frac\ri4 (\tilde\gamma^a)^{\alpha\beta}
{\cal D}_a {\cal D}^-_\alpha {\cal D}^-_\beta
-W^{-\alpha} {\cal D}^-_\alpha
+\frac14 ({\cal D}^-_\alpha W^{-\alpha})~.
\ee
Setting $q=3$ in (\ref{idDD}), we get the following equivalent form for the hypermultiplet propagator (\ref{GG})
\bea
G^{(1,1)}(1|2)=
-\int_0^\infty {\rm d}(\ri s) (\ri s)^\epsilon
\re^{\ri s \stackrel{\frown}{\square}_1}
({\cal D}^+_1)^4\bigg[
({\cal D}^-_1)^4 (u^+_1 u^+_2)
+ \Delta^{--}_1 (u^-_1 u^+_2)
\non\\
+ \stackrel{\frown}{\square}_1\frac{(u^-_1 u^+_2)^2}{(u^+_1 u^+_2)}
\bigg]
\delta^{6|8}(z_1-z_2){\mathbbm 1}~.
\label{G-mod}
\eea
Here we have applied Schwinger's proper-time representation for the inverse
d'Alembertian operator with $\epsilon$ being the ultraviolet
regularisation parameter which should be set to zero at the end of computations.
The expression (\ref{G-mod}) is manifestly covariantly analytic with respect to the first argument but the analyticity with respect to the other argument is implicit. It is an instructive exercise to check that (\ref{G-mod}) obeys ${\cal D}^+_{2\,\alpha} G^{(1,1)}(1|2) =0$.

It is natural to represent (\ref{G-mod}) as a sum of two terms
\begin{subequations}
\bea
G^{(1,1)}(1|2) &=& G_{\rm reg}^{(1,1)}(1|2) + G_{\rm sing}^{(1,1)}(1|2)~,
\label{G+G}\\
 G_{\rm reg}^{(1,1)}(1|2)&=&
-\int_0^\infty {\rm d}(\ri s) (\ri s)^\epsilon
\re^{\ri s \stackrel{\frown}{\square}_1}
({\cal D}^+_1)^4\Big[
({\cal D}^-_1)^4 (u^+_1 u^+_2)
+ \Delta^{--}_1 (u^-_1 u^+_2) \Big]\delta^{6|8}(z_1-z_2){\mathbbm 1}~,~~~~~~~~~\label{230}\\
G_{\rm sing}^{(1,1)}(1|2) &=&
-\int_0^\infty {\rm d}(\ri s) (\ri s)^\epsilon
\re^{\ri s \stackrel{\frown}{\square}_1}
({\cal D}^+_1)^4 \stackrel{\frown}{\square}_1 \frac{(u^-_1 u^+_2)^2}{(u^+_1 u^+_2)} \delta^{6|8}(z_1-z_2){\mathbbm 1}~.
\label{231}
\eea
\end{subequations}
The idea of this splitting is that, at coincident superspace points,
$G_{\rm reg}^{(1,1)}$ can have UV quantum divergences, but has no singularity
in the harmonic distribution. In contrast,
$G_{\rm sing}^{(1,1)}$ contains a harmonic singularity at coincident superspace
points which is potentially dangerous, since there is no unambiguous
procedure for regularising such divergencies. For the 4D $\cN=2$ and 5D $\cN=1$
hypermultiplet models, it was shown in \cite{KM,K-5D} that these harmonic singularities
are not dangerous because all contributions to the effective action from
the term like (\ref{231}) are vanishing owing to properties of the analytic
delta-function at coincident superspace points. However, we will show below
that the 6D distribution (\ref{231}) does give
non-vanishing contributions to the effective action which are
non-analytic and which correspond to the chiral anomaly.

The two parts of the hypermultiplet propagator (\ref{230}) and
(\ref{231}) obey the following differential equations for $\epsilon=0$
\begin{subequations}
\label{prop-G}
\bea
{\cal D}^{++}_1 G^{(1,1)}_{\rm reg}(1|2)&=& - 2(u^-_1 u^+_2)
({\cal D}^+_1)^4 \delta^{6|8}(z_1-z_2)~, \label{4.16}\\
{\cal D}^{++}_2 G^{(1,1)}_{\rm reg}(1|2) &=&0~,\label{4.17}\\
{\cal D}^{++}_1 G^{(1,1)}_{\rm sing}(1|2) &=&\delta_{\rm A}^{(3,1)}(1|2)
 +2(u^-_1 u^+_2)({\cal D}^+_1)^4 \delta^{6|8}(z_1-z_2)~,\label{4.18}\\
{\cal D}^{++}_2 G^{(1,1)}_{\rm sing}(1|2) &=& -\delta_{\rm
A}^{(1,3)}(1|2)~\label{4.19}.
\eea
\end{subequations}
The expression in the right-hand side of (\ref{4.16}) contains no
harmonic singularity and, thus, (\ref{4.16}) vanishes at
coincident superspace points because of insufficient number
of Grassmann derivatives acting on the superspace delta-function. The
expressions (\ref{4.18}) and (\ref{4.19}), in contrast, contain
harmonic singularities due to the harmonic delta-functions, see
(\ref{an-delta}).
These harmonic singularities require a regularisation which may
make the expressions (\ref{4.18}) and (\ref{4.19}) non-trivial at
coincident points.
Such singular terms should be removed from the propagator in order
to properly define  the effective current
$L^{++}_{\rm Mez}$ which has to obey the harmonic shortness
constraint (\ref{L-non-anal}).
Therefore, comparing (\ref{prop-G}) with (\ref{A+4}) and
(\ref{L-non-anal}), we conclude that at coincident superspace
points $G^{(1,1)}_{\rm reg}$ is responsible for the effective
current $L^{++}_{\rm Mez}$,
\begin{subequations}
\bea
L^{++}_{\rm Mez} = -G^{(1,1)}_{\rm reg}(1|2)|_{1=2}~,
\label{4.17a}
\eea
while $G^{(1,1)}_{\rm sing}$ generates
$\tilde L^{++}$,
\bea
\tilde L^{++} = - G^{(1,1)}_{\rm sing}(1|2)|_{1=2}~.
\label{4.17b}
\eea
\end{subequations}
While \eqref{4.17a} can be seen to contain no harmonic singularities,
a harmonic regularisation is still required to give meaning to \eqref{4.17b}.

According to eqs.\ (\ref{A+4}) and (\ref{L-non-anal}),
the anomalies $A^{(+4)}$ and
$A^{(+3)}_\alpha$ corresponding to the $V$- and $M$-formulations
of the theory are defined by the following formal expressions
\begin{subequations}
\bea
A^{(+4)} &=& -{\cal D}^{++}[G^{(1,1)}(1|2)|_{1=2}] =
 -{\cal D}^{++}[G^{(1,1)}_{\rm sing}(1|2)|_{1=2}] ~, \\
A^{(+3)}_\alpha &=& - D^+_\alpha [G^{(1,1)}_{\rm reg}(1|2)|_{1=2}]
= D^+_\alpha [G^{(1,1)}_{\rm sing}(1|2)|_{1=2}]~.
\label{4.19+}
\eea
\end{subequations}
Below, we compute the chiral anomaly $A^{(+3)}_\alpha$ by analysing
the short distance behaviour of the hypermultiplet propagator. The
structure of $A^{(+4)}$ in the abelian case will be discussed in section
\ref{Summary}.

\subsection{Covariant chiral anomaly}

According to (\ref{4.19+}), the chiral anomaly in Mezincescu's
formulation of the gauge theory is defined by the part of the
hypermultiplet propagator (\ref{231}).
Using the identity $({\cal D}^+)^4 \stackrel{\frown}{\square}
=\stackrel{\frown}{\square} ({\cal D}^+)^4$ the latter may be represented
in the equivalent form
\be
G_{\rm sing}^{(1,1)}(1|2) = \epsilon \int_0^\infty {\rm d}(\ri s)
(\ri s)^{\epsilon-1}\re^{\ri s\stackrel{\frown}{\square}_1}
\frac{(u^-_1 u^+_2)^2}{(u^+_1 u^+_2)} ({\cal D}_1^+)^4
\delta^{6|8}(z_1-z_2){\mathbbm 1}~.
\label{G-reg}
\ee
Note that this expression is analytic only in the first argument. Hence, eq.\ (\ref{4.19+})
implies
\bea
A^{(+3)}_\alpha &=& {\cal D}^+_{2\,\alpha} G^{(1,1)}_{\rm sing}
(1|2)|_{1=2} \non\\
&=& -\epsilon \int_0^\infty {\rm d}(\ri s)
(\ri s)^{\epsilon-1}\re^{\ri s\stackrel{\frown}{\square}_1}
(u^-_1 u^+_2)^2 ({\cal D}_1^+)^4 {\cal D}^-_{1\,\alpha}
\delta^{6|8}(z_1-z_2){\mathbbm 1}|_{1=2}~,
\label{G-reg1}
\eea
where we applied the identity (\ref{D-delta-identity}) to achieve
the last line.

In (\ref{G-reg1}), one has to expand the exponent of the operator
(\ref{box-anal}) in a series and accumulate eight Grassmann
derivatives to apply the identity
\be
({D}^+)^4 ({D}^-)^4 \delta^8(\theta_1-\theta_2)|_{1=2}
=1~.
\label{delta-id}
\ee
It is sufficient to consider a
covariantly constant on-shell vector multiplet
\be
{\cal D}_a W^{+\alpha}=0~,\qquad
{\cal D}^+_\alpha W^{+\alpha} = {\cal D}^-_\alpha W^{+\alpha} = 0~.
\label{on-shell-bg}
\ee
For such a background, the form of the operator (\ref{box-anal}) simplifies:
\be
\stackrel{\frown}{\square} = {\cal D}^a {\cal D}_a - W^{+\alpha} {\cal D}^-_\alpha~.
\label{simple-box}
\ee
Taking this into account, from (\ref{G-reg1}) we find
\bea
A^{(+3)}_\alpha &=&
-\frac23\epsilon \int_0^\infty
{\rm d}(\ri s)(\ri s)^{\epsilon+2}
\varepsilon_{\alpha\beta\gamma\delta}
W^{+\beta}W^{+\gamma}W^{+\delta}
\re^{\ri s {\cal D}^a {\cal D}_a}\delta^6(x_1-x_2){\mathbbm 1}|_{1=2}
\non\\&=&
-\frac\ri{96\pi^3} \varepsilon_{\alpha\beta\gamma\delta}W^{+\beta}
W^{+\gamma}W^{+\delta}
\epsilon \int_0^\infty {\rm d}(\ri s)
 (\ri s)^{\epsilon-1}~.
\eea
Here, in the last line, we applied the identity
\be
\re^{\ri\, \epsilon\,{\cal D}^a {\cal D}_a} \delta^6(x_1-x_2){\mathbbm 1}|_{x_1=x_2}
=\int\frac{{\rm d}^6 k}{(2\pi)^6}
\re^{\ri\, \epsilon\, ({\cal D}^a+\ri k^a)({\cal D}_a+\ri k_a)}{\mathbbm 1}=
-\frac{{\mathbbm 1}}{64\pi^3 \epsilon^3}+O(\epsilon^{-2})~,
\label{mom-int1}
\ee
where the terms $O(\epsilon^{-2})$ are irrelevant for small $\epsilon$.

The integration over the proper time always assumes the exponent $\re^{-\alpha s}$, $\alpha>0$, in the integrand which makes the integral convergent.
In the limit of small $\epsilon$, the identity
\be
\lim_{\epsilon\to0}\epsilon \int_0^\infty {\rm d}(\ri s)
 (\ri s)^{\epsilon-1}\re^{-\alpha s} =1
\ee
yields
\be
A^{(+3)}_\alpha =-\frac\ri{96\pi^3} \varepsilon_{\alpha\beta\gamma\delta}W^{+\beta}
W^{+\gamma}W^{+\delta}~.
\label{chir-anom}
\ee

The Lie-algebra valued field strength $W^{+\alpha}$ and gauge parameter
$\xi^{(-3)\alpha}$ may be written as linear combinations of  the generators $T^{\cal A}$
of the gauge group,
\be
W^{+\alpha} = W^{+\alpha}_{\cal A} T^{\cal A} ~,\qquad
\xi^{(-3)\alpha} = \xi^{(-3)\alpha}_{\cal A} T^{\cal A}~.
\ee
In terms of the superfields $W^{+\alpha}_{\cal A}$, the anomaly (\ref{chir-anom})
reads
\be
A_\alpha^{(+3)\cal A} = {\rm str}(T^{\cal A} A^{(+3)}_\alpha)
= -\frac\ri{96\pi^3} d^{\cal ABCD}
\varepsilon_{\alpha\beta\gamma\delta}W^{+\beta}_{\cal B} W^{+\gamma}_{\cal C} W^{+\delta}_{\cal
D}~,
\label{3.36}
\ee
where $d^{\cal ABCD}$ is the gauge-invariant tensor
(\ref{dTensorInv}). Comparing (\ref{3.36}) with (\ref{2.28}) we
find the value of the coefficient $\kappa$ for the model of
hypermultiplet in external gauge superfield
\be
\kappa = -\frac1{96\pi^3}~.
\label{kappa}
\ee

The covariant chiral anomaly manifests itself as a deformation of the analyticity
of the effective current (\ref{L-non-anal}) which is expressed through
the part of the hypermultiplet propagator (\ref{4.19})
with harmonic singularities at coincident points.
In the four-dimensional case, it was proved in \cite{KM}
that all contributions to the hypermultiplet effective action from such a term
are vanishing, and the analyticity of the effective
current is preserved.
In this paper, we demonstrate that in the six-dimensional case this term
in the hypermultiplet propagator plays an important role since it generates
the anomalous part of the hypermultiplet effective action.

\subsection{Consistency terms}

The procedure for constructing the consistent chiral anomaly in
harmonic superspace is described in sect.\ \ref{Sec2.3}. The
consistency terms are given by the non-local functional
(\ref{X-term2}). Substituting the covariant
chiral anomaly in the form (\ref{3.36}) into this functional
we get
\bea
X(y) &=&\frac{\ri}{32\pi^3} d^{\cal ABCD} y\int {\rm d}^{6|8}z_1 {\rm d}u_1 {\rm d}^{6|8}z_2 {\rm d}u_2\,
\xi^{(-3)\alpha}_{\cal A}(z_1,u_1)\varepsilon_{\alpha\beta\gamma\delta}
M^{--}_{\cal E}(z_2,u_2)
\non\\&&\times
\left(W_{\cal B}^{+\beta}(z_1,u_1)W_{\cal C}^{+\gamma}(z_1,u_1)
\frac{\delta W_{\cal D}^{+\delta}(z_1,u_1)}{\delta V_{\cal E}^{++}(z_2,u_2)}
\right)\bigg|_{V^{++}\to yV^{++}}~.
\label{X-term4}
\eea

Note that the harmonic zero-curvature equation (\ref{zero-curv})
implies the relation
\be
\frac{\delta V^{--}(z_1,u_1)}{\delta V^{++}(z_2,u_2)} = \re^{\ri b(z_1,u_1)}
({\cal D}^+_2)^4\left[
\re^{-\ri b(z_2,u_2)}
\frac{\delta^{6|8}(z_1-z_2)}{(u^+_1 u^+_2)^2}\right]~,
\label{V-V+}
\ee
where $b(z,u) = b_{\cal A}(z,u) T^{\cal A}$ is the Lie-algebra-valued bridge
superfield (see eq.\ (\ref{Dbridge})). Taking advantage of (\ref{V-V+})
and (\ref{W}), the variational derivative of the superfield
strength $W^{+\alpha}$ may be brought to the form
\be
\frac{\delta W_{\cal D}^{+\alpha}(z_1,u_1)}{\delta V_{\cal E}^{++}(z_2,u_2)} =
\frac\ri{24}\varepsilon^{\alpha\beta\gamma\delta}
{\cal D}^+_{1\beta} {\cal D}^+_{1\gamma} {\cal D}^+_{1\delta}
({\cal D}^+_2)^4\left[
\re^{\ri b(z_1,u_1)} \re^{ - \ri b(z_2,u_2)}
\frac{\delta^{6|8}(z_1-z_2)}{(u_1^+ u_2^+)^2}\right]_{\cal DE}~.
\label{varW}
\ee
Substituting this variation into (\ref{X-term4}) and integrating
over $z_2$ we get the final expression for the consistency terms
\bea
X(y) &=& \frac{d^{\cal ABCD}}{128\pi^3} y\int {\rm d}^{6|8}z {\rm d}u {\rm
d}u'\, \frac{V^{++}_{\cal E}(z,u')}{(u^+ u^+{}')^2} \non\\
&&\times
\left\{{\cal D}^+_\alpha {\cal D}^+_\beta {\cal D}^+_\gamma[\xi^{(-3)\alpha}_{\cal A}
W^{+\beta}_{\cal B} W^{+\gamma}_{\cal C}]
\left(\re^{\ri b(z,u)} \re^{ - \ri b(z,u')}\right)_{\cal DE}\right\}\Big|_{V^{++}\to
yV^{++}}~.
\label{X-final}
\eea

At the end of this section we give the resulting expression for the
consistent chiral anomaly in the $\cN= (1,0)$ superspace, which is a sum
of the covariant anomaly (\ref{Gamma-cov1}) with $A_\alpha^{(+3)\cal A}$ as in (\ref{3.36}),
 and the consistency term (\ref{X-terms-0}) with $X(y)$ given by (\ref{X-final})
\bea
\delta_\xi\Gamma &=& -\frac\ri{96\pi^3} d^{\cal ABCD} \int {\rm d}^{6|8}z {\rm d}u
\,\xi^{(-3)\alpha }_{\cal A}
\varepsilon_{\alpha\beta\gamma\delta}W^{+\beta}_{\cal B}
W^{+\gamma}_{\cal C} W^{+\delta}_{\cal D} \non\\&&
+\frac{d^{\cal ABCD}}{128\pi^3}\int_0^1{\rm d}y\, y\int {\rm d}^{6|8}z {\rm d}u {\rm
d}u'\, \frac{V^{++}_{\cal E}(z,u')}{(u^+ u^+{}')^2} \non\\
&&\times
\left\{{\cal D}^+_\alpha {\cal D}^+_\beta {\cal D}^+_\gamma[\xi^{(-3)\alpha}_{\cal A}
 W^{+\beta}_{\cal B} W^{+\gamma}_{\cal C}]
\left(\re^{\ri b(z,u)} \re^{ - \ri b(z,u')}\right)_{\cal DE}\right\}\Big|_{V^{++}\to
yV^{++}}~.
\label{final-final}
\eea
An interesting feature of this expression is that it is local in
the superspace coordinates $z^A = (x^a,\theta_i^\alpha)$, but is
non-local in the harmonics $u$. This resembles the SYM
classical action in the harmonic superspace which is also non-local in
the harmonic variables \cite{Zupnik86}.

\subsection{Abelian limit}
The field strength $W^{+\alpha}$ depends linearly on the
harmonic connection $V^{--}$, in accordance with (\ref{W}), while $V^{--}$ is a
non-linear function of the analytic prepotential $V^{++}$ in the
non-abelian case. The explicit expression for $V^{--}$ in terms of
$V^{++}$ was given in \cite{Zupnik:1987vm}.
Modulo a $\t$-gauge transformation, the bridge
superfield $b(z,u)$ is also a non-linear function of $V^{++}$
which was presented in \cite{GIOS1}. The formula (\ref{X-final})
suggests that in all these non-linear functions the analytic
prepotential $V^{++}$ should be replaced with $yV^{++}$, making
$X(y)$ a highly non-trivial function of $y$. However, in the abelian case
this dependence on $y$ simplifies such that the integration over
$y$ may be easily done in (\ref{X-terms-0}). Indeed, in the
abelian case the variational derivative (\ref{varW}) reduces to
\be
\frac{\delta W^{+\alpha}(z_1,u_1)}{\delta V^{++}(z_2,u_2)} =
\frac\ri{24}\varepsilon^{\alpha\beta\gamma\delta}
D^+_{1\beta} D^+_{1\gamma} D^+_{1\delta}
(D^+_2)^4
\frac{\delta^{6|8}(z_1-z_2)}{(u_1^+ u_2^+)^2}~,
\ee
so that the consistency term (\ref{X-term4}) reads
\bea
X(y) &=& \frac{1}{128\pi^3} y^3 \int {\rm d}^{6|8}z_1 {\rm d}u_1 {\rm
d}^{6|8}z_2{\rm d}u_2\, \xi^{(-3)\alpha}(z_1,u_1)
W^{+\beta}(z_1,u_1) W^{+\gamma}(z_1,u_1) \non\\&&\times M^{--}(z_2,u_2)
D^+_{1\beta} D^+_{1\gamma} D^+_{1\delta} (D^+_2)^4
\frac{\delta^{6|8}(z_1-z_2)}{(u^+_1 u^+_2)^2}
\non\\&=&
\frac{\ri}{32\pi^3}y^3 \int {\rm d}^{6|8}z {\rm d}u \,\xi^{(-3)\alpha}
\varepsilon_{\alpha\beta\gamma\delta} W^{+\beta} W^{+\gamma}
W^{+\delta}~.
\label{X-ab}
\eea
After integrating (\ref{X-ab}) over ${\rm d}y$ and
adding the abelian version of (\ref{3.36}), we get
\be
\delta_\xi \Gamma = -\frac\ri{384\pi^3} \int {\rm d}^{6|8}z {\rm d}u
\,\xi^{(-3)\alpha }
\varepsilon_{\alpha\beta\gamma\delta}W^{+\beta}
W^{+\gamma} W^{+\delta}~.
\ee
It shows that the consistent anomaly differs from the covariant anomaly
(\ref{chir-anom}) in the abelian limit by the factor $\frac14$.
This interplay between the coefficients in the covariant and consistent
anomalies is the same as in the non-supersymmetric case \cite{ZuminoZee,Bardeen}. This
is a non-trivial check that the result (\ref{final-final}) is the
correct expression for consistent chiral anomaly in $(1,0)$
superspace.


\section{Concluding comments}
\label{Summary}

In this paper, the chiral anomalies in
general 6D $\cN=(1,0)$ supersymmetric gauge
theories realised in harmonic superspace have been computed.
We started by recalling that there exist
two different (but related) approaches  to formulate such  gauge theories
which are based on the use of either
the analytic gauge connection $V^{++}$
or the generalised Mezincescu  prepotential $M^{--}$.
Since the gauge prepotentials  $V^{++}$ and $M^{--}$ are different
off-shell supermultiplets, the gauge transformations
in the $V$- and $M$-formulations
are also different. They are described by eqs. (\ref{deltaV}) and
\eqref{M-gauge-freedom},
respectively.
This difference implies that the chiral anomaly
manifests itself quite differently in the two formulations:
either as a deformation of the harmonic shortness
constraint of the effective current (\ref{A+4}) in the $V$-formulation
or as a deformation of the Grassmann analyticity of the effective current,
eq. (\ref{L-non-anal}), in the $M$-formulation.
In the $M$-formulation, the covariant anomaly is given by  eq. \eqref{chir-anom}.
We constructed consistency terms
such that the full chiral anomaly obeys the Wess-Zumino
consistency condition. The procedure of constructing these
consistency terms  is a generalisation of
Leutwyler's ideas \cite{Leutwyler}
to gauge theories in 6D $\cN=(1,0) $ harmonic superspace.

Our results remain valid for  the higher-derivative $\cN=(1,0) $ supersymmetric
gauge theory constructed in \cite{ISZ}. At the component level, the chiral anomalies
in this theory were discussed in \cite{Smilga}.

In this section, we will discus in some detail the issue of
constructing the chiral anomaly in the $V$-formulation of gauge
theory. In particular, we deduce an expression for the
chiral anomaly $A^{(+4)}$ in the abelian case. To construct this
anomaly, we will follow the procedure proposed in our recent work \cite{KNS}
which allows one to restore $A^{(+4)}$ when the expression
for $A^{(+3)}_\alpha$ is known.

In the $M$-formulation, the gauge theory is described by the
effective current $L^{++}_{\rm Mez}(z,u)= u^+_i u^+_j L^{ij}_{\rm
Mez}(z)$ obeying
\be
D^+_\alpha L^{++}_{\rm Mez} =\ri\, \kappa\, \varepsilon_{\alpha\beta\gamma\delta}
W^{+\beta} W^{+\gamma} W^{+\delta}~,
\label{DL++}
\ee
where $\kappa$ is given by (\ref{kappa}) for the model of
hypermultiplet interacting with background vector multiplet.
Let us introduce a superfield $F^{++}(z,u)$ as a solution of the
equation
\be
D^+_\alpha F^{++} = \ri \,\kappa\,\varepsilon_{\alpha\beta\gamma\delta}
W^{+\beta} W^{+\gamma} W^{+\delta}
\label{F-eq}
\ee
and defined modulo arbitrary shift of the form
\be
F^{++} \rightarrow F^{++}+ H^{++}~,\qquad
D^+_\alpha H^{++} =0~.
\label{Freedom}
\ee
A particular solution of (\ref{F-eq}) is
\be
F^{++} = -\frac\ri2\, \kappa\, V_{\alpha\beta} W^{+\alpha} W^{+\beta}
-\frac\ri{64}\, \kappa\, \varepsilon^{\alpha\beta\gamma\delta} V_{\alpha\beta}
V_{\gamma\delta} D^+ W^+~,
\label{F++}
\ee
where $V_{\alpha\beta}$ is the connection defined in (\ref{Vab}).
This solution has the following important property
\be
D^{++} F^{++} = -\frac\ri2 \kappa\, G^{++\alpha\beta}
\partial_{\alpha\beta} V^{++}~,
\label{DF++}
\ee
where
\be
G^{++\alpha\beta} = W^{+\alpha} W^{+\beta} + \frac1{16}
\varepsilon^{\alpha\beta\gamma\delta} V_{\gamma\delta} D^+ W^+~,
\qquad
D^+_\gamma G^{++\alpha\beta} =0~.
\ee
The property (\ref{DF++}) shows that $D^{++} F^{++}$ is analytic
and, thus, may appear as a part of the anomaly superfield
$A^{(+4)}$ in the $V$-formulation of the gauge theory.

Let us now introduce the following superfield
\be
{\mathbb L}^{++} = L^{++}_{\rm Mez} - F^{++}~,
\ee
which is analytic due to the properties (\ref{DL++}) and (\ref{F-eq}),
\be
D^+_\alpha {\mathbb L}^{++} = 0~.
\ee
However, unlike $L^{++}_{\rm Mez}$, this superfield is no longer
holomorphic on ${\mathbb C}P^1$,
\be
D^{++} {\mathbb L}^{++} = {\mathbb A}^{(+4)} ~, \qquad
D^+_\alpha {\mathbb A}^{(+4)} =0~.
\ee
Thus, the chiral anomaly is completely encoded in the analytic
superfield ${\mathbb A}^{(+4)}$ which is defined modulo shifts
\be
{\mathbb A}^{(+4)} \rightarrow {\mathbb A}^{(+4)}  - D^{++}
H^{++}~,
\label{AAH}
\ee
which follow from the freedom in the definition of $F^{++}$, see
(\ref{Freedom}).

For the choice of $F^{++}$ as in eq.\ (\ref{F++}), the superfield
$ {\mathbb A}^{(+4)}$ reads
\be
{\mathbb A}^{(+4)} = \frac \ri2 \kappa\, G^{++\alpha\beta}
\partial_{\alpha\beta} V^{++}~.
\label{AA4}
\ee
However, this superfield is not yet the anomaly $A^{(+4)}$ which
corresponds to the $V$-formulation of the gauge theory with effective
current $L^{++}_{\rm an}$ obeying (\ref{A+4}). The problem is that
(\ref{AA4}) does not satisfy the Wess-Zumino consistency
condition. Indeed, it varies under the gauge transofrmation
(\ref{deltaV}) as
\be
\delta_\lambda {\mathbb A}^{(+4)} = -\frac\ri2 \kappa\,
D^{++}(G^{++\alpha\beta} \partial_{\alpha\beta}\lambda)~.
\ee
This implies that ${\mathbb A}^{(+4)} $
 cannot appear as the  gauge variation of an
effective action, $\delta_\lambda\Gamma=\int {\rm d}\zeta^{(-4)}\lambda
{\mathbb A}^{(+4)}$, since
$\delta_{\lambda_1}\delta_{\lambda_2} \G \neq
\delta_{\lambda_2}\delta_{\lambda_1} \Gamma $.

To resolve this problem,
the arbitrariness (\ref{AAH}) should be employed. Indeed, we propose the
following non-local expression for $H^{++}$
\begin{subequations}
\label{HB}
\bea
H^{++}(\zeta) &=& -\int {\rm d}\zeta'^{(-4)} G^{(2,0)}(\zeta,\zeta')
{\mathbb B}^{(+4)}(\zeta')~,
\label{HB1}\\
{\mathbb B }^{(+4)} &=& \frac\ri2 \kappa\, \partial_{\alpha\beta}(V^{++}
G^{++\alpha\beta})~,
\label{HB2}
\eea
\end{subequations}
where $G^{(2,0)}(\zeta,\zeta')$ is Green's function
\be
G^{(2,0)}(\zeta,\zeta') = \frac1\square
(D^+)^4 (D^+{}')^4\left[
\delta^{6|8}(z-z')\frac{(u^+ u^-{}')}{(u^+ u^+{}')^3}
\right]
\ee
with the property
\be
D^{++} G^{(2,0)}(\zeta,\zeta') = \delta_{\rm
A}^{(4,0)}(\zeta,\zeta')~.
\ee
The expression (\ref{HB}) is chosen such that the superfield
\bea
A^{(+4)} &=& {\mathbb A}^{(+4)} - D^{++} H^{++} =
{\mathbb A}^{(+4)} + {\mathbb B}^{(+4)}
\non\\
&=& \ri\,\kappa\, G^{++\alpha\beta} \partial_{\alpha\beta} V^{++}
+\frac\ri2\kappa\, V^{++} \partial_{\alpha\beta} G^{++\alpha\beta}
\label{A4-res}
\eea
is analytic, $D^+_\a A^{(+4)} =0$,
 and obeys the Wess-Zumino consistency condition.
Indeed,
we can consistently associate with (\ref{A4-res}) an effective action $\G$, since
the variation defined by
\be
\delta_\lambda\Gamma = \int {\rm d}\zeta^{(-4)} \lambda A^{(+4)}
\ee
is integrable,
\be
(\delta_{\lambda_1}\delta_{\lambda_2} -
\delta_{\lambda_2}\delta_{\lambda_1})
\Gamma =0~.
\label{int-cond}
\ee
We stress that the integrability condition (\ref{int-cond}) is
nontrivial already in the abelian case since the function
(\ref{A4-res}) is not gauge invariant. This confirms that this
function can consistently describe the chiral anomaly in the
$V$-formulation of the gauge theory.

It is very tempting to derive $A^{(+4)} $
by direct  supergraph computations in the harmonic
superspace.\footnote{It would be also very interesting to study the problem
of computing the chiral anomalies in $\cN=(1,0)$ supersymmetric gauge theories
using the supergraph technique in projective superspace
 developed in \cite{Davgadorj:2017ezp} and references therein.}
Another important issue is to construct a
generalisation of (\ref{A4-res}) to the case of non-abelian chiral
anomaly. We leave these issues for future studies.\\


\noindent
{\bf Acknowledgements:}\\
We thank Ian McArthur for comments on the manuscript.
IBS is grateful to Evgeny Buchbinder and Ian McArthur for useful discussions,
and to the School of Physics and Astrophysics at UWA,
where the major part of this work was done, for kind hospitality.
IBS acknowledges the support from the RFBR grant No
15-02-06670. SMK and IBS were supported in part by the Australian Research Council,
project No. DP140103925.
The work of SMK is also supported in part by the Australian Research Council,
project No. DP160103633. JN acknowledges support
from GIF the German-Israeli Foundation for Scientific Research and Development.

\appendix

\section{Vector multiplet in harmonic superspace}

Supersymmetric Yang-Mills theory in 6D
$\cN=(1,0)$ harmonic superspace was formulated
in \cite{HSWest,Zupnik86,Zupnik:1987vm}.
Here we briefly review this formulation
following the harmonic superspace notation of \cite{GIOS}.

Let $u^+_i$ and $u^-_i$ be  standard $\sSU(2)$ harmonic variables,
$(u_i^-\, ,u_i^+) \in \sSU(2)$,
\be
\overline{u^{+i}} = u^-_i~,
\qquad u^{+i}u_i^- = 1~,
\ee
with $ u^+_i = \ve_{ij}u^{+j}$.
Let $D^{++}$, $D^{--}$ and  $D^0$ be the associated
harmonic derivatives defined as in \cite{GIOS}.
Using the harmonics we introduce
a new basis for the gauge-covariant spinor derivatives
\be
{\cal D}^\pm_\alpha = u^\pm_i {\cal D}^i_\alpha
= D^\pm_\alpha + \ri\, V^{\pm}_\alpha~,\qquad
V^\pm_\alpha = u^\pm_i V^i_\alpha~.
\label{cov-D}
\ee
In accordance with (\ref{gauged-algebra}) the operators
(\ref{cov-D}) obey the following (anti)commutation relations
\begin{subequations}
\label{algebra}
\bea
\{ {\cal D}^+_\alpha , {\cal D}^+_\beta  \} &=&0~,\label{B5a}\\
\{  {\cal D}^+_\alpha , {\cal D}^-_\beta \} &=& 2\,\ri\,
(\gamma^a)_{\alpha\beta} {\cal D}_a~,\label{B5b}\\
{}[{\cal D}_a , {\cal D}^\pm_\alpha] &=& \ri\, (\gamma_a)_{\alpha\beta} W^{\pm \beta}~,\label{B5c}\\
{}[{\cal D}_a, {\cal D}_b] &=& \ri\, F_{ab}~,
\label{B5d}
\eea
\end{subequations}
where $W^{\pm\alpha}$ are the irreducible $\sU(1)$ components  of the
field strength $W^{i\alpha}$,
\bea
W^{\pm \alpha} = u^\pm_i W^{i\alpha}\,.
\label{Harm-proj}
\eea

In the harmonic superspace setting, it is useful to combine the superspace
gauge-covariant derivatives with the harmonic ones,
\be
{\cal D}_{\hat A} = ({\cal D}_a , {\cal D}^\pm_\alpha,
{\cal D}^{++},{\cal D}^{--}, {\cal D}^0)
:=({\cal D}_a , {\cal D}^\pm_\alpha,D^{++},D^{--}, D^0)
= {D}_{\hat A} +\ri\, V_{\hat A}
~.
\label{tau-frame}
\ee
The gauge transformation of $\cD_{\hat A} $
is analogous to (\ref{2.2}),
\be
{\cal D}_{\hat A} \longrightarrow
{\cal D}'_{\hat A}  = \re^{\ri \t}{\cal D}_{\hat A} \re^{-\ri \t}~.
\ee
Since the gauge superfield parameter $\tau$ is harmonic independent, the harmonic derivatives
$(D^{\pm\pm},D^0)$ are gauge covariant.

The equation (\ref{B5a}) is the integrability condition for
covariantly analytic superfields to exist.
This equation can be solved in terms of a bridge superfield $b = b(z,u)$
 defined by the rule
\be
{\cal D}^+_\alpha = \re^{-\ri b } D^+_\alpha \re^{\ri b}~.
\label{Dbridge}
\ee
The introduction of the bridge superfield leads to a new gauge freedom, in addition to the $\tau$-gauge transformations (\ref{2.2}). The complete gauge transformation
law of $b$ is
\be
\re^{\ri b'} = \re^{\ri\lambda} \re^{\ri b}\re^{-\ri\tau}~,
\ee
where $\lambda$ is a $\sU(1)$ neutral analytic superfield,
$D^+_\alpha\lambda =0$.

The representation (\ref{tau-frame}) for the gauge-covariant derivatives is called the $\tau$-frame. When it is important, we will attach a label `$(\tau)$' to the covariant derivatives in this representation, ${\cal D}_{(\tau)\hat A}$. Using the bridge superfield one can introduce another representation for these derivatives, which is usually referred to as the $\lambda$-frame,
\be
{\cal D}_{(\tau)\hat A} \longrightarrow
{\cal D}_{(\lambda)\hat A} = \re^{\ri b}{\cal D}_{(\tau)\hat A} \re^{-\ri b}
= {D}_{\hat A} +\ri\, V_{(\lambda)\hat A}~.
\ee
Below, we will consider all relations in the $\lambda$-frame, and we will omit the label `$(\lambda)$'.
In the $\lambda$-frame, the derivative ${\cal D}^+_\alpha$ is short,
${\cal D}^+_{\alpha} = D^+_\alpha$, and hence $V_{\a}^+=0$.
However,  two of the three harmonic derivatives acquire gauge connections:
\be
{\cal D}^{++} = D^{++} + \ri \,V^{++}~,\qquad
{\cal D}^{--} = D^{--} + \ri\, V^{--}~.
\ee
As follows from the commutation relation $[{\cal D}^+_\alpha , {\cal D}^{++}]=0$, the gauge connection $V^{++}$ is analytic,
\be
D^+_\alpha V^{++} = 0~.
\ee
The connection  $V^{--}$
 can be expressed via $V^{++}$ as a unique solution of
 the zero-curvature condition
\be
[{\cal D}^{++}, {\cal D}^{--}] = D^0 \quad
\Longleftrightarrow \quad
D^{++} V^{--} - D^{--} V^{++} + \ri\, [V^{++}, V^{--}] =0~.
\label{zero-curv}
\ee
The explicit expression for $V^{--}$ in terms of $V^{++}$ was originally found
by Zupnik \cite{Zupnik:1987vm}.
In the $\lambda$-frame, no $\t$-gauge freedom remains.
Under the $\lambda$-gauge group,
the connections $V^{++}$ and $V^{--}$ transform as
\bea
V'^{\pm\pm}=\re^{{\rm i}\lambda}V^{\pm\pm}\re^{-{\rm i}\lambda}
-\ri \,\re^{{\rm i}\lambda}D^{\pm\pm}\re^{-{\rm i}\lambda}~.
\label{A.122}
\eea

The $\l$-frame counterparts of
the (anti-)commutation relations (\ref{B5b}) and \eqref{B5c},
in conjunction with the identity $[{\cal D}^{--}, {\cal D}^+_\a] = {\cal D}^-_\a$,
allow one to express the gauge connections
 $V^-_\a$ and $V_a$ and the field strength  $W^{+\alpha}$
in terms of $V^{--}$. The explicit expressions for the connections are
\begin{subequations}
\bea
V^-_\alpha &=& - D^+_\alpha V^{--}~, \\
\qquad
V_a &=& \frac\ri8 (\tilde\gamma_a)^{\alpha\beta}
D^+_\alpha D^+_\beta V^{--} \quad
\Longleftrightarrow \quad
V_{\a\b} = (\gamma^a)_{\alpha\beta} V_a =
\frac\ri2 D^+_\alpha D^+_\beta V^{--}~.~~~~~~~
\label{Vab}
\eea
\end{subequations}
The expression for the  field strength  is
\be
W^{+\alpha} = \frac\ri{24} \varepsilon^{\alpha\beta\gamma\delta}
D^+_\beta D^+_\gamma D^+_\delta V^{--}~.
\label{W}
\ee
As mentioned above, $V^{--}$ is uniquely expressed
in terms of the analytic connection $V^{++}$.
Thus, the superfield $V^{++}$ is a single  prepotential in terms of which all the connections are determined, in complete analogy with the
4D case \cite{GIKOS,Zupnik:1987vm}.
This prepotential is analytic, but otherwise unconstrained.


\begin{footnotesize}

\end{footnotesize}


\begin{thebibliography}{66}

\bibitem{WZ-cons}
J.~Wess and B.~Zumino,
``Consequences of anomalous Ward identities,''
  Phys.\ Lett.\ B {\bf 37}, 95 (1971).


\bibitem{Fujikawa} K.~Fujikawa,
``Path integral measure for gauge invariant fermion theories,'' Phys.~Rev.~Lett. {\bf 42}, 1195 (1979);
  ``Path integral for gauge theories with fermions,''
  Phys.\ Rev.\ D {\bf 21}, 2848 (1980)
  Erratum: [Phys.\ Rev.\ D {\bf 22}, 1499 (1980)].

\bibitem{Stora}
R. Stora, ``Algebraic structure and topological origin of anomalies,''
in:  {\it Recent Progress in Gauge Theories},
G 't Hooft, A. Jaffe, H. Lehmann, P. K. Mitter, I. M. Singer and R. Stora
(Eds.), Plenum Press, New York, 1984, pp. 543--559.

\bibitem{Zumino}
B. Zumino, ``Chiral anomalies and differential geometry,''
 in:  {\it Relativity, Groups and
Topology II}, B. S. De Witt and R. Stora (Eds.),
  North Holland, Amsterdam, 1984, pp. 1291--1322.

\bibitem{A-GW}
  L.~Alvarez-Gaume and E.~Witten,
  ``Gravitational anomalies,''
  Nucl.\ Phys.\ B {\bf 234}, 269 (1984).

\bibitem{ZuminoZee}
  B.~Zumino, Y.~S.~Wu and A.~Zee,
  ``Chiral anomalies, higher dimensions, and differential geometry,''
  Nucl.\ Phys.\ B {\bf 239}, 477 (1984).

\bibitem{Bardeen}
  W.~A.~Bardeen and B.~Zumino,
  ``Consistent and covariant anomalies in gauge and gravitational theories,''
  Nucl.\ Phys.\ B {\bf 244}, 421 (1984).

\bibitem{Leutwyler}
  H.~Leutwyler,
  ``Chiral fermion determinants and their anomalies,''
  Phys.\ Lett.\ B {\bf 152}, 78 (1985).

\bibitem{FS}
  K.~Fujikawa and H.~Suzuki,
 {\it Path Integrals and Quantum Anomalies},
  Oxford University Press, Oxford, 2004, 284 p.

\bibitem{BvN}
  F.~Bastianelli and P.~van Nieuwenhuizen,
 {\it Path Integrals and Anomalies in Curved Space},
 Cambridge University Press, Cambridge, 2006, 398 p.

\bibitem{TS}
  P.~K.~Townsend and G.~Sierra,
  ``Chiral anomalies and constraints on the gauge group in higher-dimensional supersymmetric {Yang-Mills} theories,''
  Nucl.\ Phys.\ B {\bf 222}, 493 (1983).

\bibitem{FK1}
P.~H.~Frampton and T.~W.~Kephart,
``Explicit evaluation of anomalies in higher dimensions,''
  Phys.\ Rev.\ Lett.\  {\bf 50}, 1343 (1983)
  Erratum: [Phys.\ Rev.\ Lett.\  {\bf 51}, 232 (1983)].

\bibitem{FK2}
  P.~H.~Frampton and T.~W.~Kephart,
  ``Consistency conditions for Kaluza-Klein axial anomalies,''
  Phys.\ Rev.\ Lett.\  {\bf 50}, 1347 (1983).

 \bibitem{FK3}
P.~H.~Frampton and T.~W.~Kephart,
``Analysis of anomalies in higher space-time dimensions,''
Phys.\ Rev.\ D {\bf 28}, 1010 (1983).

\bibitem{PS84}
  O.~Piguet and K.~Sibold,
  ``The anomaly in the Slavnov identity for N=1 supersymmetric Yang-Mills theories,''
  Nucl.\ Phys.\ B {\bf 247}, 484 (1984).

\bibitem{CL84}
  T.~E.~Clark and S.~T.~Love,
  ``Supersymmetric effective actions for anomalous internal chiral symmetries,''
  Phys.\ Lett.\  {\bf 138B}, 289 (1984).

\bibitem{Nielsen84}
  N.~K.~Nielsen,
  ``Anomalies of supersymmetric chiral Yang-Mills currents,''
  Nucl.\ Phys.\ B {\bf 244}, 499 (1984).

\bibitem{GGS}
  G.~Girardi, R.~Grimm and R.~Stora,
  ``Chiral anomalies in $N=1$ supersymmetric {Yang-Mills} theories,''
  Phys.\ Lett.\  {\bf 156B}, 203 (1985).

\bibitem{GKM85}
  E.~Guadagnini, K.~Konishi and M.~Mintchev,
  ``Non-abelian chiral anomalies in supersymmetric gauge theories,''
  Phys.\ Lett.\  {\bf 157B}, 37 (1985).

\bibitem{KS85}
  K.~i.~Konishi and K.~i.~Shizuya,
  ``Functional integral approach to chiral anomalies in supersymmetric gauge theories,''
  Nuovo Cim.\ A {\bf 90}, 111 (1985).

\bibitem{BPT1}
  L.~Bonora, P.~Pasti and M.~Tonin,
  ``ABJ anomalies in supersymmetric Yang-Mills theories,''
  Phys.\ Lett.\  {\bf 156B}, 341 (1985).

\bibitem{BPT2}
  L.~Bonora, P.~Pasti and M.~Tonin,
  ``The consistent chiral anomaly in supersymmetric Yang-Mills theories,''
  Nucl.\ Phys.\ B {\bf 261}, 249 (1985)
  Erratum: [Nucl.\ Phys.\ B {\bf 269}, 745 (1986)].

\bibitem{OMc}
  I.~N.~McArthur and H.~Osborn,
  ``Supersymmetric chiral effective action and nonabelian anomalies,''
  Nucl.\ Phys.\ B {\bf 268}, 573 (1986).

\bibitem{J}
  Y.~Ohshima, K.~Okuyama, H.~Suzuki and H.~Yasuta,
  ``Remark on the consistent gauge anomaly in supersymmetric theories,''
  Phys.\ Lett.\ B {\bf 457}, 291 (1999)
  [hep-th/9904096].

\bibitem{Maja}
  M.~Marinkovic,
  ``Wess-Zumino effective action for supersymmetric Yang-Mills theories,''
  Nucl.\ Phys.\ B {\bf 366}, 74 (1991).

 \bibitem{GGKPS}
  S.~J.~Gates  Jr., M.~T.~Grisaru, M.~E.~Knutt, S.~Penati and H.~Suzuki,
  ``Supersymmetric gauge anomaly with general homotopic paths,''
  Nucl.\ Phys.\ B {\bf 596}, 315 (2001)
  [hep-th/0009192].

\bibitem{FWZ}
S.~Ferrara, J.~Wess and B.~Zumino,
``Supergauge multiplets and superfields,''
  Phys.\ Lett.\ B {\bf 51}, 239 (1974).

\bibitem{BSohnius}
P.~Breitenlohner and M.~F.~Sohnius,
``Superfields, auxiliary fields, and tensor calculus for N=2 extended supergravity,''
Nucl.\ Phys.\  B {\bf 165}, 483 (1980).

\bibitem{KNS}
  S.~M.~Kuzenko, J.~Novak and I.~B.~Samsonov,
  ``The anomalous current multiplet in 6D minimal supersymmetry,''
  JHEP {\bf 1602}, 132 (2016) [arXiv:1511.06582 [hep-th]].

\bibitem{HSezgin}
  P.~S.~Howe and E.~Sezgin,
  ``Anomaly free tensor Yang-Mills system and its dual formulation,''
  Phys.\ Lett.\ B {\bf 440}, 50 (1998)
  [hep-th/9806050].

\bibitem{HST}
  P.~S.~Howe, G.~Sierra and P.~K.~Townsend,
  ``Supersymmetry in six dimensions,''
  Nucl.\ Phys.\ B {\bf 221}, 331 (1983).

\bibitem{Koller}
  J.~Koller,
  ``A six-dimensional superspace approach to extended superfields,''
  Nucl.\ Phys.\ B {\bf 222}, 319 (1983).


\bibitem{Siegel}
  W.~Siegel,
  ``Superfields in higher dimensional space-time,''
  Phys.\ Lett.\ B {\bf 80}, 220 (1979).

\bibitem{Nilsson}
  B.~E.~W.~Nilsson,
  ``Superspace action for a 6-dimensional non-extended supersymmetric
  Yang-Mills theory,''
  Nucl.\ Phys.\ B {\bf 174}, 335 (1980).


\bibitem{BP}
  I.~L.~Buchbinder and N.~G.~Pletnev,
  ``Construction of 6D supersymmetric field models in N=(1,0) harmonic superspace,''
  Nucl.\ Phys.\ B {\bf 892}, 21 (2015)
  [arXiv:1411.1848 [hep-th]].

\bibitem{BP2}
  I.~L.~Buchbinder and N.~G.~Pletnev,
  ``Leading low-energy effective action in the 6D hypermultiplet theory on a vector/tensor background,''
  Phys.\ Lett.\ B {\bf 744}, 125 (2015)
  [arXiv:1502.03257 [hep-th]].

\bibitem{BP3}
  I.~L.~Buchbinder, B.~S.~Merzlikin and N.~G.~Pletnev,
  ``Induced low-energy effective action in the 6D, $\mathcal{N} =(1,0)$ hypermultiplet theory on the vector multiplet background,''
  Phys.\ Lett.\ B {\bf 759}, 626 (2016)
  [arXiv:1604.06186 [hep-th]].

\bibitem{BP4}
  I.~L.~Buchbinder, E.~A.~Ivanov, B.~S.~Merzlikin and K.~V.~Stepanyantz,
  ``One-loop divergences in the $6D, \mathcal N = (1,0)$ abelian gauge theory,''
  Phys.\ Lett.\ B {\bf 763}, 375 (2016)
  [arXiv:1609.00975 [hep-th]].

\bibitem{BP5}
  I.~L.~Buchbinder, E.~A.~Ivanov, B.~S.~Merzlikin and K.~V.~Stepanyantz,
  ``One-loop divergences in 6D, $ \mathcal{N} $ = (1, 0) SYM theory,''
  JHEP {\bf 1701}, 128 (2017)
  [arXiv:1612.03190 [hep-th]].

\bibitem{BP6}
I.~L.~Buchbinder, E.~A.~Ivanov, B.~S.~Merzlikin and K.~V.~Stepanyantz,
  ``Supergraph analysis of the one-loop divergences in $6D$,
  ${\cal N} = (1,0)$ and ${\cal N} = (1,1)$ gauge theories,''
  Nucl.\ Phys.\ B {\bf 921}, 127 (2017)
  [arXiv:1704.02530 [hep-th]].

\bibitem{HSWest}
  P.~S.~Howe, K.~S.~Stelle and P.~C.~West,
  ``$N=1$, $d = 6$ harmonic superspace''
  Class.\ Quant.\ Grav.\  {\bf 2}, 815 (1985).

\bibitem{Zupnik86}
  B.~M.~Zupnik,
  ``Six-dimensional supergauge theories in the harmonic superspace,''
  Sov.\ J.\ Nucl.\ Phys.\  {\bf 44}, 512 (1986)
  [Yad.\ Fiz.\  {\bf 44}, 794 (1986)].



\bibitem{Zupnik:1987vm}
  B.~M.~Zupnik,
  ``The action of the supersymmetric $N=2$ gauge theory in harmonic superspace,''
  Phys.\ Lett.\ B {\bf 183}, 175 (1987).


\bibitem{GIKOS}
  A.~Galperin, E.~Ivanov, S.~Kalitzin, V.~Ogievetsky and E.~Sokatchev,
  ``Unconstrained $N=2$ matter, Yang-Mills and supergravity theories in harmonic superspace,''
  Class.\ Quant.\ Grav.\  {\bf 1}, 469 (1984).



\bibitem{GIOS}
A.~S.~Galperin, E.~A.~Ivanov, V.~I.~Ogievetsky and E.~S.~Sokatchev,
{\it Harmonic Superspace}, Cambridge University Press,  2001.



\bibitem{Mezincescu}
  L.~Mezincescu,
  ``On the superfield formulation of O(2) supersymmetry,''
 Dubna preprint JINR-P2-12572 (June, 1979).


\bibitem{BFM}
  I.~L.~Buchbinder, E.~I.~Buchbinder, S.~M.~Kuzenko and B.~A.~Ovrut,
  ``The background field method for N=2 super-Yang-Mills theories in harmonic superspace,''
  Phys.\ Lett.\ B {\bf 417}, 61 (1998)
  [hep-th/9704214].


\bibitem{CommentsBFM}
  I.~L.~Buchbinder and S.~M.~Kuzenko,
  ``Comments on the background field method in harmonic superspace: Non-holomorphic corrections in N=4 SYM,''
  Mod.\ Phys.\ Lett.\ A {\bf 13}, 1623 (1998)
  [hep-th/9804168].

\bibitem{KM0}
  S.~M.~Kuzenko and I.~N.~McArthur,
  ``Effective action of N=4 super Yang-Mills: N=2 superspace approach,''
  Phys.\ Lett.\ B {\bf 506}, 140 (2001)  [hep-th/0101127].


\bibitem{KM}
  S.~M.~Kuzenko and I.~N.~McArthur,
  ``Hypermultiplet effective action: N = 2 superspace approach,''
  Phys.\ Lett.\ B {\bf 513}, 213 (2001)
  [hep-th/0105121].


\bibitem{K-5D}
  S.~M.~Kuzenko,
  ``Five-dimensional supersymmetric Chern-Simons action as a hypermultiplet quantum correction,''
  Phys.\ Lett.\ B {\bf 644}, 88 (2007)
  [hep-th/0609078].


\bibitem{LT-M}
 W.~D.~Linch III and G.~Tartaglino-Mazzucchelli,
  ``Six-dimensional supergravity and projective superfields,''
JHEP {\bf 1208}, 075 (2012)
[arXiv:1204.4195 [hep-th]].
%

\bibitem{BPT87}
  L.~Bonora, P.~Pasti and M.~Tonin,
  ``Chiral anomalies in higher dimensional supersymmetric theories,''
  Nucl.\ Phys.\ B {\bf 286}, 150 (1987).


\bibitem{deWG}
  B.~de Wit and M.~T.~Grisaru,
  ``Compensating fields and anomalies,''
 in  {\it Quantum Field Theory and Quantum Statistics, Vol. 2},
 I. A.  Batalin, C. J. Isham and G. A. Vilkovisky (Eds.) Adam Hilger, Bristol,
 1987, pp. 411--432.

\bibitem{KL}
  S.~M.~Kuzenko and W.~D.~Linch,
  ``On five-dimensional superspaces,''
  JHEP {\bf 0602}, 038 (2006) [hep-th/0507176].


\bibitem{GIOS1}
  A.~Galperin, E.~Ivanov, V.~Ogievetsky and E.~Sokatchev,
  ``Harmonic supergraphs: Green functions,''
  Class.\ Quant.\ Grav.\  {\bf 2}, 601 (1985).


  \bibitem{ISZ}
  E.~A.~Ivanov, A.~V.~Smilga and B.~M.~Zupnik,
  ``Renormalizable supersymmetric gauge theory in six dimensions,''
  Nucl.\ Phys.\ B {\bf 726}, 131 (2005)
  [hep-th/0505082].

\bibitem{Smilga}
  A.~V.~Smilga,
  ``Chiral anomalies in higher-derivative supersymmetric 6D theories,''
  Phys.\ Lett.\ B {\bf 647}, 298 (2007)
  [hep-th/0606139].

\bibitem{Davgadorj:2017ezp}
  A.~Davgadorj and R.~von Unge,
  ``$\mathcal{N} = 2$ super Yang-Mills theory in projective superspace,''
  arXiv:1706.07000 [hep-th].

\end{thebibliography}
\end{document}